# Flash annealing-engineered wafer-scale relaxor antiferroelectrics for enhanced energy storage performance

Yizhuo Li[1]†, Kepeng Song[3]†, Meixiong Zhu[1]†, Xiaoqi Li[1,2], Zhaowei Zeng[1,2], KangMing Luo[1,2], Yuxuan Jiang[1,2], Zhe Zhang[1,2], Cuihong Li[1], Yujia Wang[1,2], Bing Li[1,2], Zhihong Wang[4], Zhidong Zhang[1], and Weijin Hu[1,2]*

[1]Shenyang National Laboratory for Materials Science, Institute of Metal Research, Chinese Academy of Sciences, Shenyang 110016, People's Republic of China.

[2]School of Materials Science and Engineering, University of Science and Technology of China, Shenyang 110016, People's Republic of China.

[3]Electron Microscopy Center, School of Chemistry & Chemistry Engineering, Shandong University, Jinan 250100, People's Republic of China.

[4]Electronic Materials Research Laboratory, School of Electronic Science and Engineering, Xi'an Jiaotong University, Xi'an 710049, People's Republic of China.

* Corresponding email: wjhu@imr.ac.cn

† These authors contributed equally.

**Dielectric capacitors are essential for energy storage systems due to their high-power density and fast operation speed. However, optimizing energy storage density with concurrent thermal stability remains a substantial challenge. Here, we develop a flash annealing process with ultrafast heating and cooling rates of 1000 °C/s, which facilitates the rapid crystallization of $PbZrO_3$ film within a mere second, while locking its high-temperature microstructure to room temperature. This produces compact films with sub-grain boundaries fraction of 36%, nanodomains of several nanometers, and negligible lead volatilization. These contribute to relaxor antiferroelectric film with a high breakdown strength (4800 kV/cm) and large polarization (70 μC/cm²). Consequently, we have achieved a high energy storage density of 63.5 J/cm³ and outstanding thermal stability with performance degradation less than 3% up to 250 °C. Our approach is extendable to ferroelectrics like $Pb(Zr_{0.52}Ti_{0.48})O_3$ and on wafer scale, providing on-chip nonlinear dielectric energy storage solutions with industrial scalability.**

**One-Sentence Summary:** Flash annealing ultrafast-crystallizes $PbZrO_3$, yielding a relaxor antiferroelectric state that boosts energy storage performance.

## INTRODUCTION

Dielectric capacitors store electrical energy through electric displacement, owning distinct merits including high power density and ultrafast charge/discharge speed, are promising for applications in modern energy storage devices such as consumer electronics, high-power lasers, and radar systems (*1-3*). Nevertheless, compared to chemical energy sources like Li-ion batteries and solid oxide fuel cells, they possess a relatively low energy density (*4*). Consequently, extensive efforts have been devoted to enhance the energy density of dielectric materials with high efficiency, in response to the ongoing trend of device miniaturization in electronic systems (*5-7*). The device performance is evaluated by the recoverable energy density $U_e = \int_{P_r}^{P_m} E dP$ and the energy storage efficiency $\eta = U_e/(U_e + U_{loss})$, where $P_m$ and $P_r$ are the maximum and remanent polarization, respectively (**Fig. S1**). $U_{loss}$ is determined by the hysteresis of $P$-$E$ loop, which not only limits $U_e$, but also degrades $\eta$, resulting in energy dissipation problem that deteriorates the device thermal stability. Hence, a large $P_m$ with a minor hysteresis is vital for achieving a good energy storage performance. In this regard, nonlinear dielectrics such as ferroelectrics (FE) and antiferroelectrics (AFE) are competitive because of large $P_m$. However, they usually suffer from apparent loss associated with the switching of the FE domains or the first-order field-induced AFE-FE phase transition (*8*).

To address this issue, the main idea has been to develop relaxor FE (RFE) or relaxor AFE (RAFE) by reducing the domain size from micrometer to nanometer scale, which can weaken the domain intercoupling and break the long-range-ordered domains into short-range-ordered nanodomains (*9*). This process lowers the switching barrier and smears out the hysteresis, thereby enhancing both $U_e$ and $\eta$. Based on this framework, approaches including multi-phase composition (*10*), solid solution (*11*), chemical doping (*5*), and defect engineering (*12*) have been put forth to facilitate the design of RFE or RAFE materials. Representative examples include constructing room temperature RAFE in $AgNbO_3$ via solid solution with $AgTaO_3$ to obtain a $U_e$ of 6.3 J/cm$^3$ and $\eta$ of 90% (*11*); creating a superparaelectric state in $BiFeO_3$-$BaTiO_3$ RFE system by chemical doping and get a $U_e$ of 152 J/cm$^3$ and $\eta$ of 77% (*5*); or achieving a $U_e$ of 133 J/cm$^3$ and $\eta$ of 75% in 0.68Pb$(Mg_{1/3}Nb_{2/3})O_3$-0.32$PbTiO_3$ RFE film by introducing point defects through high-energy ion bombardment (*12*). In addition to the nanodomain design, grain refinement has also been explored to augment energy storage capabilities, recognizing that the reduction in grain size ($d$) can enhance the breakdown field ($E_b \propto d^{-\alpha}$, where $\alpha \sim$ 0.2–0.4) across diverse dielectrics (*13, 14*). Implementing methods include chemical doping (*15*), multi-phase composition (*16*), high-entropy stabilization (*17*), and hot-pressing treatment (*18*). However, these approaches are commonly associated with several drawbacks: the requirement for exact stoichiometry and precise control of interface microstructures necessitates intricate fabrication procedures, which not only elevate production costs but also complicate manufacturing processes, creating substantial barriers to large-scale production. To overcome these challenges, there is an urgent need for in-depth research and technological innovation.

We noticed that FEs or AFEs display relaxor characteristics with a large amount of nanodomains near Curie temperature ($T_c$) (**Fig. 1A**), because of the flattened domain-switching pathway with reduced energy barrier at high temperature. If one can freeze the high-temperature nanodomains down to room temperature (**Fig. 1A**), a slim *P-E* loop with a large $P_m$ is expected to sustain, leading to large $U_e$ and $\eta$. Inspired by the conventional quenching techniques employed in structural materials (*19*), we believe an intuitive approach to achieve this goal involves rapid crystallization succeeded by quenching, which tends to inhibit the normal grain growth and the emergence of long-range ordered domains, thereby retaining the nanodomains even at room temperature. To conduct this process, we develop a flash heating and cooling method (FHC) that features both high heating and cooling rate (~1000 °C/s), enabling the ultrafast sintering within less than 1 second. Specifically, we put the films together with a graphite substrate into a copper coil that can rapidly heat the film through electromagnetic induction and thermal conduction (**Fig. 1B**). Additionally, the non-contact heating feature allows us to execute the flash cooling within 1 second by immersing the films into the liquid nitrogen (LN2).

## RESULTS

**FHC-engineered relaxor antiferroelectric characters.** To test the feasibility of FHC, we synthesized $PbZrO_3$ (PZO) AFE films, which are extensively considered for energy storage applications. We commenced by spin-coating PZO amorphous films via chemical solution deposition, followed by FHC treatment (**Fig. 1C, Table S1**). For comparison, we conducted a series of control treatments, including conventional muffle furnace annealing (CA), rapid thermal annealing (RTA), and flash heating with subsequent air cooling (FH). The heating rates differ substantially, with 1 °C/s for CA, 30 °C/s for RTA, and a remarkable 1000 °C/s for both FH and FHC, enabling a thorough examination of heating rates on material properties over three orders of magnitude. The cooling process involves air cooling for CA, RTA within a time of ~ 1000 s, FH within 30 s, and FHC within a mere 1 s (**Fig. S2**). **Fig. 1D** illustrates that PZO film produced by FHC process exhibits a slim *P-E* loop with negligible hysteresis, a characteristic distinct from the typical AFE double-hysteresis loops for films fabricated by other methods. The corresponding switching current curves (*J-E,* inset of **Fig. 1D**) reveal four peaks which are smaller in intensity and broaden in width for the FHC film, demonstrating its relaxor antiferroelectric (RAFE) characteristics. We also performed similar FHC process by varying the heating rates (**Fig. S3**). Substantial relaxor behavior emerges when the heating rate is up to ~ 500 °C/s. These findings suggest that both flash heating (i.e., within 650 ms) and LN2 quenching (i.e., < 1 s) are critical for developing the RAFE state. The evolution from AFE to RAFE has also been reflected in the dielectric ($\varepsilon_r$) spectrum (**Figs. 1E & 1F**). $\varepsilon_r$ at 1 kHz and ambient temperature are 167, 168, 206, and 404 for films processed by CA, RTA, FH, and FHC, respectively, with the RAFE-FHC film achieving a 2.5-fold enhancement in $\varepsilon_r$ than that of AFE film. This increase in $\varepsilon_r$ aligns well with **Fig. 1A**, wherein the RAFE phase exhibits smaller domains, making them easier to be rotated by the electric field. With increasing the temperature, AFE treated by CA, RTA, and FH exhibit a typical sharp dielectric peak at $T_c$ ~ 220-240 ºC (see also **Fig. S4**). Conversely, FHC film exhibits a pronounced broadening of $\varepsilon_r$ over a wide temperature range, indicating a profound diffuse phase transition between RAFE and paraelectric states. The relaxor diffuseness factor $\gamma$, as derived from the modified Curie-Weiss law (*20*), is ~ 2.0 for FHC film, notably higher than AFE phase treated by other methods ($\gamma$ = 1.5, 1.3, 1.7 for CA, RTA, and FH films) (**Fig. S5**). Interestingly, the FHC induced relaxor behavior exhibits a minimal frequency dispersion, with $T_c$ differing by only ~3 K between 1 kHz and 1 MHz (**Fig. 1F**). This $T_c$ shift is smaller than other PZO relaxor AFEs engineered by chemical doping such as $Li^+$–$Al^{3+}$ co-doping (~ 10 K) (*21*), and by $Al_2O_3$ interface engineering (~ 6 K) (*22*). Consistently, the dielectric loss (tan $\delta$, **Fig. S6**) also exhibits minimal frequency dispersion, maintaining small values between 0.01 and 0.05 across all frequencies. This exceptionally weak frequency dependence is attributed to the paraelectric-like crystal structure and robust grain boundaries produced by FHC process, as detailed in subsequent sections. Moreover, the relaxor *P-E* loops of FHC film remain stable from 25 ºC to 250 ºC (**Fig. 1H**), demonstrating a remarkable thermal stability. In sharp contrast, *P-E* loops of AFE film gradually transformed into the RAFE state with increasing temperature, showing strong thermal dependence (**Fig. 1G**). This contrast is further evidenced in current switching curves (**Fig. S7**), where FHC films display a smaller AFE-FE transition field that is weakly dependent on temperature. Such a weak frequency and temperature dispersion of $\varepsilon_r$ and *P-E* loop for FHC film suggests its suitability for applications requiring stringent frequency and thermal stability, including high-precision capacitors and energy storage devices.

**The microscopic origin of FHC-induced relaxor antiferroelectrics.** To gain insights into the structure origin of this robust RAFE state, we performed high-resolution X-ray diffraction (HRXRD, **Fig. S8**) on PZO films by using synchrotron X-ray source at 10 keV. While all films display similar diffraction patterns, there is a systematic shift of diffraction peaks to lower angles with increasing the heating rates from CA to RTA, and further to FH and FHC, implying the expansion in lattice constants and cell volume. Consequently, we chose the AFE-RTA and RAFE-FHC films to investigate their temperature-dependent HRXRD profiles, as depicted in **Figs. 2A & 2B**. As indicated by the dashed arrows, the diffraction peaks of AFE-RTA film shift towards lower angles as the temperature rises, whereas, they remain unchanged for RAFE-FHC film. Additionally, the expanded view of the diffraction profile around the (200) reflection (**Fig. 2C**) and at other planes-including (100), (110), and (211) (**Fig. S9**)-reveals distinct peak splitting (~0.4° at 25 °C) in the AFE-RTA film, attributed to tetragonal distortion. In contrast, no peak splitting appears in RAFE-FHC film. This structure difference has been quantitatively reflected in their lattice constant values (**Fig. 2D**) as derived from HRXRD. Notably, the lattice constants *a* & *c* of AFE film (Tetragonal, space group P4mm) are clearly different at room temperature, giving rise to a *c*/*a* ratio of ~ 0.9893. The sharp increase of *c* near $T_c$ of ~ 230 ºC makes them eventually merge into the paraelectric state (Cubic, space group Pm-3m). Consequently, the RTA film exhibits notable anisotropic thermal expansion in its AFE state. Such anisotropy, has also been observed in PZO bulk ceramics (*23*), is characteristic of materials with reduced crystal symmetry. Meanwhile, the lattice constants *a* & *c* of RAFE film are nearly equal and identical to those of the paraelectric phase. As a result, RAFE film exhibits a minor *c*/*a* ratio expansion (from 0.999 to 1), in sharp contrast with that of AFE film (from 0.989 to 1) across $T_c$ (inset of **Fig. 2D**). This comparison clearly suggests that FHC treatment can effectively preserve the lattice configuration of the high-temperature state down to room temperature. The paraelectric-like RAFE state with minor structural distortion enables the formation of nanodomain with size shrinking down to 1-3 nm, as evidenced by the High-Angle Annular Dark-Field Scanning Transmission Electron Microscopy (HAADF-STEM) image presented in **Fig. 2F**. In contrast, AFE-RTA film exhibits long range-ordered domains with large size of more than ~10 nm (**Fig. 2E**). This strong sensitivity of AFE/RAFE configuration on the structure distortion *c*/*a* has been further revealed by the phase-field simulation (**Figs. 2G & 2H**). We found that the free energy barrier between the non-polar AFE state and PE states reduces substantially from $7.20\times10^7$ J/m$^3$ to $0.02\times10^7$ J/m$^3$ as the *c*/*a* ratio increases from 0.989 to 1.000, leading to the transformation of long-range ordered domain to the disordered nanodomain with reduced domain size (**Fig. 2H**). The decreased barrier facilitates the phase transition, thereby reducing the field required to achieve the saturation polarization and effectively minimizing energy losses in the hysteresis loops. Correspondingly, the double-hysteresis *P-E* loops change into the slim relaxor loops as the *c*/*a* increases (**Fig. S10**), a change that aligns well with the experimental observations (**Fig. 1D**).

In addition to the nanodomain engineering, FHC also have prominent impact on the grain distributions, as revealed by surface images shown in **Figs. 3A & 3B, and Fig. S11.** Take RTA-AFE (**Fig. 3A**) and FHC-RAFE (**Fig. 3B**) films as an example, both share similar grain size distribution with average grain size (AGS) of 183 and 197 nm, respectively. However, RAFE film has a rather smooth surface with a small average roughness $R_a$ of ~0.65 nm compared to that of RTA-AFE film ($R_a$ ~ 1.28 nm), because of the denser packed crystal grains characterized by a narrow grain boundary width (GBW) and shallow GB depth (GBD). The average GBW & GBD are ~ 22 nm & 1.0 nm for FHC-RAFE film, whereas they increase to ~ 68 nm & ~ 8 nm for RTA-AFE film (**Figs. S11D & E**). This noticeable difference is also clearly visible in the TEM images (**Fig. S12**). We represent these metric parameters in **Fig. 3C**, highlighting that both the FH and FHC processes substantially improve the grain compactness and surface smoothness, thanks to the ultrafast crystallization that effectively avoids the grain contraction during the cooling process. To reveal more details of GB, we further performed Transmission Kikuchi Diffraction (TKD), and obtained the Inverse Pole Figure (IPF, **Figs. 3D & 3E**) mapping for FHC-RAFE and RTA-AFE films. The IPF analysis reveals that RAFE film has more sub-grain boundaries (sub-GB, indicated by white dashed circles) characterized by misorientation angle of 1-2$^\circ$ than that of AFE film. As summarized in **Fig. 3F**, the fraction of sub-GB increases from 10.1% for AFE film to 35.9% for RAFE film. In contrast, the normal large-angle GB (misorientation angle > 15$^\circ$) decreases from 87.7% for AFE film to 56.7% for RAFE film. A typical sub-GB, marked by the dashed line, is further visualized by the STEM image (**Fig. 3G**), which is associated to a perfect dislocation with a Burgers vector of **a**[101]**.** By performing Fast Fourier Transform (FFT) on the STEM image, we observed two sets of barely distinguishable diffraction spots (inset of **Fig. 3G**), and the misorientation angle of this sub-GB is determined to be ~1.7°. Moreover, we identified a characteristic GB with a larger misorientation angle of ~4.5°, which is associated with a network of dislocations (**Fig. S13**). This finding is consistent with the consensus that the formation of GB/sub-GB is a consequence of dislocation pile-up at the boundary (*24*). We attribute the formation of sub-GB to the FHC process, where flash cooling within 1 s suppresses grain growth and raises the dislocation density within grains, promoting sub-GBs formation via the rearrangement and intersections of dislocations during the cooling process. The cumulative effect of crystal misalignment and misfit dislocation at sub-GBs impedes long-range domain ordering, promoting nanodomain formation and stabilizing the RAFE phase. Concurrently, sub-GBs induce notable internal stress, as qualified by Kernel Average Misorientation (KAM) maps (**Fig. S14**) showing larger intra-grain misorientation angles at GBs and sub-GBs. This confirms intensified lattice distortion and residual stress, that can be further quantified via Williasson-Hall (W-H) analysis (*25*) of XRD peak broadening (**Fig. S15**). Residual stress increases from 0.036% for CA, 0.13% for RTA, 0.12% for FH, and ultimately to 0.18% for FHC, demonstrating how thermal processing rates critically tailor PZO film microstructures to achieve RAFE properties.

Furthermore, FHC effectively suppresses Pb volatilization, a common issue in conventional long-time annealing processes. This suppression is corroborated by EDS analysis (**Fig. 3H**), which shows a systematic increase in Pb content with reduced processing time: the Pb: Zr atomic ratio rises from 0.84±0.07 (CA) to 1.01±0.08 (FHC). Additional EDS mapping (**Fig. S16**) further indicates that Pb deficiency primarily localizes at grain boundaries. This Pb deficiency is likely the root cause of the subtle dielectric kink observed in CA films (**Fig. 1E**). The stoichiometric Pb: Zr ratio in RAFE films is further supported by photoluminescence (PL) spectra (**Fig. 3I**). The characteristic PL peak attributed to Pb vacancies ($V_{Pb}$), observed at ~0.5 eV away from the band edge in RTA film, is absent in the RAFE-FHC film. The position of this PL peak aligns well with energy levels of $V_{Pb}$ predicted from first-principles calculations (~0.7 eV, **Fig. S17**). Concurrently, the bandgap of RAFE film (~3.06 eV) is ~0.3 eV smaller than that of AFE film (~3.37 eV), which is attributed to the combined effect of lattice constants and Pb vacancy concentrations, as supported by the bandgap values summarized in **Table S2**.

**Electric properties and energy storage performance.** To evaluate the energy-storage performance, we measured *P-E* loops by applying large electric field (**Fig. 4A**). Compared to other films, RAFE-FHC film sustains a larger electric field while maintains a big $P_m$ of ~ 70 μC/cm$^2$. As revealed by the *P-E* loops under increasing electric fields (**Fig. S18**), this large polarization originates from the field-induced RAFE-FE transition. The Weibull distribution plots (**Fig. 4B**) show that the RAFE-FHC film has an excellent breakdown electric field ($E_b$) of 4827 kV/cm, much higher than that of AFE films treated by other processes (CA: 1228 kV/cm; RTA: 1826 kV/cm; FH: 2638 kV/cm). Meanwhile, the Weibull modulus $\beta$ increases from 9.8 for AFE-CA film to 49.4 for RAFE-FHC film, suggesting the improved reliability and uniformity by FHC process. It is well known that there is an intrinsic trade-off between $\varepsilon_r$ and $E_b$, with a larger $\varepsilon_r$ usually signifies a lower $E_b$ for dielectric materials (**Fig. 4C**) (*14, 26, 27*). Interestingly, our RAFE-PZO film has both good $\varepsilon_r$ and $E_b$. We ascribe this substantial improvement in $E_b$ to the FHC induced compact grains, sub-grain boundaries, and pronounced suppression of Pb vacancies. These features efficiently block leakage paths, enhance the electron scattering, and reduce the carrier concentrations, drastically diminishing the leakage current to ~ $10^{-8}$ A/cm$^2$ under a dc field of 400 kV/cm for the RAFE-FHC film—five orders of magnitude smaller than that of the AFE-CA film (~$2\times10^{-3}$ A/cm$^2$) (**Fig. 4D)**.

The remarkable combination of high $E_b$, large $P_m$, and slender hysteresis result in a high $U_e$ of 63.5 J/cm$^3$ for RAFE-FHC film, which is 2-4 times that of AFE films treated by CA (~ 17.3 J/cm$^3$), RTA (~ 20.1 J/cm$^3$), and FH (~ 27.9 J/cm$^3$) (**Fig. 4E**). We also found a maximum energy storage efficiency $\eta$ of ~ 80% at an $E$ of < 2000 kV/cm, which remains at level of ~ 60% at $E_b$ for the RAFE-FHC film. As revealed in **Fig. S19**, this substantial performance enhancement primarily stems from microstructure engineering (sub-grains and nano-domains), with Pb stoichiometry providing secondary gains. Notably, the energy storage performance maintains over a broad temperature range up to 250 °C (**Fig. 4F**). This excellent thermal stability guarantees proper functioning of the device under extreme temperature conditions, including the hybrid electric vehicles (<140 °C) and the harsh environments of underground oil/gas exploration (<200 °C) (*28, 29*). We ascribe it to the stable relaxor behavior of RAFE state (**Fig. 1H**). The FHC process effectively preserves the high-temperature structure, endowing the film with temperature-insensitive characteristics. Additionally, the excellent insulating nature with minor $V_{Pb}$ defects mitigate the thermal activation of carriers, thereby minimizing the conduction losses at elevated temperatures (**Fig. 1E**). Finally, both RTA and FHC films maintain stable energy storage performance up to $10^7$ cycles without breakdown (**Fig. S20**). This endurance is comparable to state-of-the-art values reported for thin-film capacitors based on PZO (~$10^7$) (*30*) and other materials such as $BiFeO_3$-$BaTiO_3$ solid solutions (~$10^8$) (*5*). We summarize the metrics of pristine PZO films fabricated using diverse techniques including magnetic sputtering, pulsed laser deposition (PLD), chemical solution deposition (CSD) processed with microwave radiation, among others, in **Table I** (*30-34*). Our RAFE-FHC capacitor device boasts the best performance, featuring exceptional thermal stability with $U_e$ and $\eta$ degradation of less than 3% up to 250 °C. The flash annealing treatment, requiring less than 1 second, is at least two orders of magnitude faster than other methods, suggesting its high efficiency in producing RAFE-PZO films.

Additionally, the minor temperature-dependent lattice variation of the FHC-RAFE film, enabling it to withstand extreme thermal cycling test from LN2 to 400 °C, as shown in **Fig. 5A**. After 100 thermal cycles, the *P-E* loop of RTA-AFE capacitor broadens substantially (**Fig. 5B**), while the FHC-RAFE capacitor exhibits only subtle changes (**Fig. 5C**). This is consistent with the optical images (inset of **Figs. 5B & 5C)**: the RTA-AFE film has more pores and structural defects after cycling, whereas the FHC-RAFE film maintains its structural integrity. Consequently, the FHC-RAFE capacitor experiences minor degradation with energy storage density dropping from 63.5 J/cm³ to 63.3 J/cm³ and efficiency from 60.1% to 56.9%, much smaller than that of RTA-AFE capacitor with energy density declining from 20.1 J/cm³ to 12.3 J/cm³ and efficiency dropping from 69% to 38.7% (**Fig. 5D**). Importantly, the FHC treatment is compatible with large-scale production. As represented in inset of **Fig. 5E**, we fabricated a RAFE film on a 2-inch platinized silicon wafer. XRD results from 9 typical positions show lattice constants *a* & *c* consistent with the RAFE state and a minor variation within 0.08% (**Fig. 5E**), indicating exceptional structural homogeneity. The film also achieves uniform energy storage characteristics, with an $U_e$ of 54.5 ± 2.7 J/cm³ and an $\eta$ of 69.1 ± 4.9% (**Fig. 5F**). The excellent tolerance to extreme thermal cycling renders FHC-RAFE highly suitable for applications involving frequent temperature changes, such as in the aerospace field, electric vehicles, and 5G communication base stations where capacitors encounter complex thermal environments. Moreover, the results underscore the FHC's advantage in high-throughput manufacturing of RAFE films and micro-capacitors, providing on-chip energy storage solutions with industrial scalability.

## DISCUSSION

We have developed a flash annealing technique characterized by ultrafast heating and cooling rates reaching 1000 °C/s, which enables the rapid crystallization of $PbZrO_3$ films within a mere second, effectively locking the high-temperature nanodomain configurations down to room temperature, giving rise to a relaxor antiferroelectric state with enhanced energy storage performance including the energy storage density and the excellent thermal stability. Note that this ultrafast crystallization process is also compatible with Pt as the top electrode, demonstrated by RAFE like *P-E* loop of a typical $La_{0.67}Sr_{0.33}MnO_3$/PZO/Pt capacitor treated by FHC process (**Fig. S21**), confirming viability for device integration. The nanodomain and sub-grain design by FHC can also be used on typical FEs like $Pb(Zr_{0.52}Ti_{0.48})O_3$ (PZT), transforming it into the RFE state with $U_e$ being enhanced by over 5-fold from 10.7 J/cm³ to 57.8 J/cm³, while keeping $\eta$ at above 62.3% (**Fig. S22**). This demonstrates its versatility in promoting the relaxor domain state and the energy storage performance of nonlinear dielectrics. More generally, the strategy is applicable to systems where nanodomain and sub-grain engineering are critical in determining the specifical functionalities, such as dielectric and piezoelectric materials.

## MATERIALS AND METHODS

**Amorphous thin-films fabrication.** We grew amorphous PZO thin films on $Pt/Ti/SiO_2/Si$ substrates via chemical solution deposition (CSD). First, we dissolved stoichiometric $Pb(CH_3COO)_2{\cdot}3H_2O$ in 2-methoxyethanol at 120 °C for 240 min, adding a 10% lead excess to compensate for annealing loss. After cooling to room temperature, we introduced $Zr(OCH_2CH_2CH_3)_4$ and stirred the solution for 60 min. We then adjusted the concentration to 0.4 mol/L by diluting with 2-methoxyethanol, matching our prior work (35). After aging the solution for 24 hours, we spin-coated it onto substrates at 3000 rpm for 30 s and pyrolyzed it at 450 °C for 5 min, and repeated this cycle to achieve the desired thickness (**Fig. S23**). To prevent final lead loss, we applied a PbO capping layer via spin coating and then performed various thermal treatments.

**Setup for the flash heating (FH) and flash heating and cooling (FHC) processes.** We assembled the electromagnetic induction heating equipment with an induction unit (Haidewei-HDWG), program control system, sample stage, and infrared pyrometer (INPTEK-IS30). We designed the sample stage as a 30-mm-diameter high-purity graphite disk placed in a quartz crucible and suspended a liquid nitrogen nozzle above it. The super high-frequency induction unit allows current adjustment (0-10 A) and frequency tuning (50-400 kHz). We coupled the infrared pyrometer (±0.5% reading +2 °C accuracy, 5 ms response) to an electromagnetic valve at the liquid nitrogen outlet, enabling millisecond quenching upon reaching target temperatures. Our system controls heating rates up to a maximum value of 1500°C/s and subsequent temperature holds or cooling in air (FH) or liquid nitrogen (FHC)."

**The conventional annealing (CA) and rapid thermal annealing (RTA) processes.** For CA, we heated the amorphous PZO film in a muffle furnace to 700 °C at 1 °C/s, held it for 30 min, then cooled it in air to room temperature. For RTA, we rapidly heated the film to 650 °C at 30 °C/s in an RTA furnace, held it for 180 s, then cooled it in air to room temperature."

**Structural characterizations.** We investigated the phase composition and stress changes in PZO thin films at various temperatures using high-resolution XRD (HRXRD) with a 10 keV synchrotron source at Beamline BL02U2 of the Shanghai Synchrotron Radiation Facility (SSRF). We analyzed atomic structures using probe-corrected STEM (Thermo Fisher Spectra 300, 200 kV) equipped with a Gatan Continuum 1065 detector, setting HAADF collection angles to 50–200 mrad and convergence angles to 21.6 mrad. We prepared cross-sectional samples via FIB (Thermo Fisher Helios G4), achieving 60 pm spatial resolution. To identify local dipoles, we tracked $Pb^{2+}$ displacement vectors (strong contrast) relative to four nearest $Zr^{4+}$ centers (weak contrast) and performed 2D Gaussian fitting using custom MATLAB code (36).

We characterized surface morphologies by atomic force microscopy (AFM, Bruker Multimode-8) in tapping mode using Pt/Ir-coated Si cantilevers, and analyzed grain size distributions via the intercept method (Heyn method, Nano Measurer). This method calculates grain dimensions as the total intercept length of random test lines divided by the number of GB intersections (**Fig. S24**). We quantified GB width/depth using NanoScope Analysis, performing line scans across 100 randomly selected GBs in AFM images. We determined grain orientations/microstructures via Transmission Kikuchi Diffraction (TKD) using a scanning electron microscope (Gemini SEM 300) outfitted with a EBSD detector (Oxford Symmetry S2), and analyzed data using Aztec Crystal 2.1. We measured composition using energy-dispersive X-ray spectroscopy at 10kV (EDS, Oxford Instruments Ultim Max 100). We recorded the photoluminescence spectrums using a fluorometer (Hitachi F-4500) with a 250 nm excitation wavelength.

**Device fabrication and electric characterizations.** For electrical measurements, we fabricated capacitor devices with Pt top electrodes (50 × 50 μm$^2$) deposited by a dc magnetron sputtering (JGP560CIV) and patterned by a standard photolithography and lift-off processes (URE-2000/35). We performed *P-E* loops, dielectric spectrums, and leakage current using a FE tester (Precision Multiferroic, Radiant Tech), an LCR meter (Tonghui TH2838), and a multisource meter (Keithley 2450) equipped on a high temperature probe station (Gogo Instruments Tec., LRT-001-D4).

**Weibull distribution Analysis.** We calculated the value of breakdown electric field ($E_b$) via the following formula (*37*):

$$X_i = \ln E_i \tag{1}$$

$$Y_i = \ln\ln\frac{1}{1-P_i} \tag{2}$$

$$P_i = \frac{i}{n+1} \tag{3}$$

where $n$ = 10 represents the number of devices, $E_i$ denotes the electric breakdown strength of each sample arranged in ascending order ($E_1 \leq E_2 \leq E_3 ... \leq E_n$), $P_i$ represents the probability of dielectric breakdown. By utilizing the two-parameter Weibull distribution function, we established a linear relationship between $X_i$ and $Y_i$, and then extracted the average $E_b$ from the point where the fitting line intersects with the horizontal axis at $Y_i = 0$. Subsequently, we obtained $P_s$, $P_r$, and hysteresis loss values at each sample's breakdown electric field.

**Phase-field simulations.** In the 3D phase-field modelling of the PZO system, the three components of the polarization serve as order parameters. The temporal evolution of polarization vector $P_i$ is governed by the time-dependent Ginzburg-Landau equation (*38*):

$$\frac{\partial P_i(\mathbf{r},t)}{\partial t} = -L_i\frac{\delta F}{\delta P_i(\mathbf{r},t)} + \xi_i(\mathbf{r},t), (i = 1,2,3) \tag{4}$$

where $L$ assumes the role of the kinetic coefficient associated with the domain wall mobility and $\delta F/\delta P_i(\mathbf{r},t)$ signifies the driving force for the spatial and temporal evolution of $P_i$. $\xi_i(r,t)$ represents the Gaussian random fluctuation, satisfying the conditions $\langle\xi_i(r,t)\rangle = 0$ and $\langle\xi_i(r,t)\xi_j(r',t')\rangle = 2k_BTL_i\delta_{ij}\delta(r-r')\delta(t-t')$, where $k_B$ is the Boltzmann constant. The total free energy $F$ encompasses contributions from bulk, gradient, elastic, and electrostatic energies, formulated as:

$$F = \iiint(f_{bulk} + f_{grad} + f_{elas} + f_{elec})dV \tag{5}$$

where $V$ is the volume of the film. The bulk free energy density is expressed as $f_{bulk} = \alpha_{ij}P_iP_j + \alpha_{ijkl}P_iP_jP_kP_l + \alpha_{ijklmn}P_iP_jP_kP_lP_mP_n$, where $\alpha_i$, $\alpha_{ij}$, *and* $\alpha_{ijk}$ are Landau coefficients. The expression for the gradient energy is given as $f_{grad} =$ $-\gamma_{11}\theta_i^2 \sum_i P_{i,i}^2 - \gamma_{12}\theta_i^2 \sum_{i\neq j\neq k} \left(P_{j,k}^2 + P_{k,j}^2\right) + g_{11} \sum_i \left(\frac{\partial^2 P_i}{\partial x_i^2}\right)^2 + g_{12} \sum_{i\neq j} \left(\frac{\partial^2 P_i}{\partial x_j^2}\right)^2$. The first two terms, describing the coupling between the oxygen tilt ($\theta$) and polarization, characterize the tendency of neighboring polarizations to align antiparallel. The oxygen tilt is estimated as $\theta = \sqrt{-\frac{k_0}{b_0}(T - T_\theta)}$, where $T_\theta$ is the transition temperature of the oxygen tilt. $k_0$ and $b_0$ are positive material constants representing the coefficients of the quadratic and quartic terms in the potential energy expression for the oxygen tilt $F_\theta = \sum_i \left(\frac{k_0(T-T_\theta)}{2}\theta_i^2 + \frac{b_0}{4}\theta_i^4\right)$. The last two terms, accounting for next-nearest-neighbor interactions, play a critical role in driving the AFE-FE phase transition (*39, 40*). The elastic energy density can be written as $f_{elas} = \frac{1}{2} C_{ijkl}\left(\varepsilon_{ij} - \varepsilon_{ij}^0\right)\left(\varepsilon_{kl} - \varepsilon_{kl}^0\right)$, in which $\varepsilon_{ij}$ is the total strain and $\varepsilon_{ij}^0$ is the spontaneous strain. The spontaneous strain is related to be the polarization through electrostrictive coefficients $Q_{ijkl}$: $\varepsilon_{ij}^0 = Q_{ijkl}P_kP_l$. The last term is the electrostatic energy density: $f_{elec} = -\frac{1}{2}\varepsilon_0\varepsilon_b E_i^2 - E_iP_i$, where $\varepsilon_0$ is the permittivity of vacuum and $\varepsilon_b$ is the background dielectric constant (*41*). The material parameters of PZO are sourced from prior literatures (*39, 40*): the Landau parameters are given by $\alpha_1 = 2.61\times10^5(T - T_0)$, where $T_0$ is the Curie temperature (510 K), $\alpha_{11} = 5.60 \times 10^5$ J m$^5$ C$^{-4}$, $\alpha_{12} = 2.89 \times 10^8$ J m$^5$ C$^{-4}$, $\alpha_{111} = 1.65 \times 10^9$ J m$^9$ C$^{-6}$, $\alpha_{112} = -8.66 \times 10^8$ J m$^9$ C$^{-6}$, and $\alpha_{123} = 3.19 \times 10^{10}$ J m$^9$ C$^{-6}$. The elastic constants are $C_{11} = 15.6 \times 10^{10}$ N m$^{-2}$, $C_{12} = 9.6 \times 10^{10}$ N m$^{-2}$, and $C_{44} = 12.7 \times 10^{10}$ N m$^{-2}$. The electrostrictive coefficients are $Q_{11} = 0.048$ m$^4$ C$^{-2}$, $Q_{12} = -0.015$ m$^4$ C$^{-2}$, and $Q_{44} = 0.047$ m$^4$ C$^{-2}$. To effectively capture the experimentally observed temperature-dependent tetragonality of AFE PZO, we linearly fitted the temperature dependent coefficients of the first-order gradient energy $\gamma_{11}\theta_i^2$ and $\gamma_{12}\theta_i^2$, yielding $\gamma_{11}\theta_i^2 = \gamma_{12}\theta_i^2 = 12.94a_c^4 \times 10^7$ J m C$^{-2}$ at 300 K and $\gamma_{11}\theta_i^2 = \gamma_{12}\theta_i^2 = 8.65a_c^4 \times 10^7$ J m C$^{-2}$ at 500 K, while the second-order gradient coefficient was fixed at $g_{11} = g_{12} = 3.32a_c^4 \times 10^7$ J m C$^{-2}$. $a_c$ is the size of the mesh grid, 0.416 nm.

We implemented the phase-field equations via the finite element method (FEM) using the open-source FERRET package (*42*) built on the Multiphysics Object-Oriented Simulation Environment (MOOSE) framework (*43*). The simulation size is $64\Delta x \times 64\Delta x \times 2\Delta x$ with a finite element mesh spacing of $\Delta x = 0.416$ nm, corresponding to the cubic lattice constant of PZO. We applied periodic boundary conditions in-plane and imposed stress-free conditions on the displacement field out-of-plane while fixing the electrostatic potential at $\Phi = 0$. For the coupled variable system, we employed third-order Hermite finite elements. We solved the resulting nonlinear algebraic equations with the Newton-Raphson method, using the generalized minimal residual method (GMRES) to handle the linear systems. We address time-dependent evolution using the implicit Euler method. The initial condition for all simulations was a paraelectric state with a small, randomly distributed noise.

**First-principles calculations.** We performed first-principles calculations within the framework of the density functional theory (DFT), as implemented in the Vienna Ab initio Simulation Package (VASP) code (*44*). We described interactions between ions and valence electrons using the projector-augmented wave (PAW) formalism (*45*) and treated the exchange-correlation functional using the generalized gradient approximation (GGA) in the Perdew-Burke-Ernzerhof (PBE) formulation (*46*). We employed a 540 eV energy cutoff for the plane-wave basis and explicitly treated Pb( 6*s*6*p*), Zr(4*s*4*p*4*d*5*s*), and O(2*s*2*p*) as valence states. For the AFE phase (40 atoms, Pbam space group), we used a 6×3×4 k-point mesh. To model Pb volatilization, we removed one Pb atom from 8-unit-cell superstructure. We fully relaxed all crystal structures with convergence thresholds of $10^{-6}$ eV for energy and 0.01 eV/Å for forces. While PBE underestimates band gaps, comparative analysis of different configurations remains valid.

**Acknowledgments:** We thank BL02U2 beamline of Shanghai Synchrotron Radiation Facility for providing beamtime, and thank Neng He and Bo Gao for their valuable discussion. **Funding:** This work was supported by the National Natural Science Foundation of China (NSFC) (Grant Nos. 52402133, 61974147, 52031014, 52122101, 52471022, 92477120), National Key R&D Program of China (Grant No. 2022YFA1203903), International Partnership Program of Chinese Academy of Sciences (CAS) (Grant No. 172GJHZ2024044MI), and Special Fund for Central Government Guiding the Local Development of Science and Technology (Grant No. 2023JH6/100100063). Y. J. W. acknowledges the Youth Innovation Promotion Association CAS (2021187). **Author contributions:** Conceptualization: W.J.H. Methodology: Y.Z.L, W.J.H., and Y.J.W. Investigation: Y.Z.L, K.P.S, M.X.Z, X.Q.L, Z.W.Z, K.M.L, Y.X.J, C.H.L, Y.J.W, B.L, and Z.H.W. Visualization: Y.Z.L, K.P.S, M.X.Z, K.M.L, Y.J.W, and W.J.H. Supervision: W.J.H. Writing – original draft: Y.Z.L, and W.J.H. Writing – review & editing: Y.Z.L, W.J.H, K.P.S, Y.J.W, and Z.D.Z. **Competing interests:** The authors declare no competing interests. A provisional patent application, titled “A device for induction heat treatment of dielectric materials”, has been applied for through Institute of Metal Research, Chinese Academy of Sciences (China provisional patent 202411658275.4). **Data and materials availability:** All data needed to evaluate the conclusions in the paper are present in the paper and/or the Supplementary Materials.

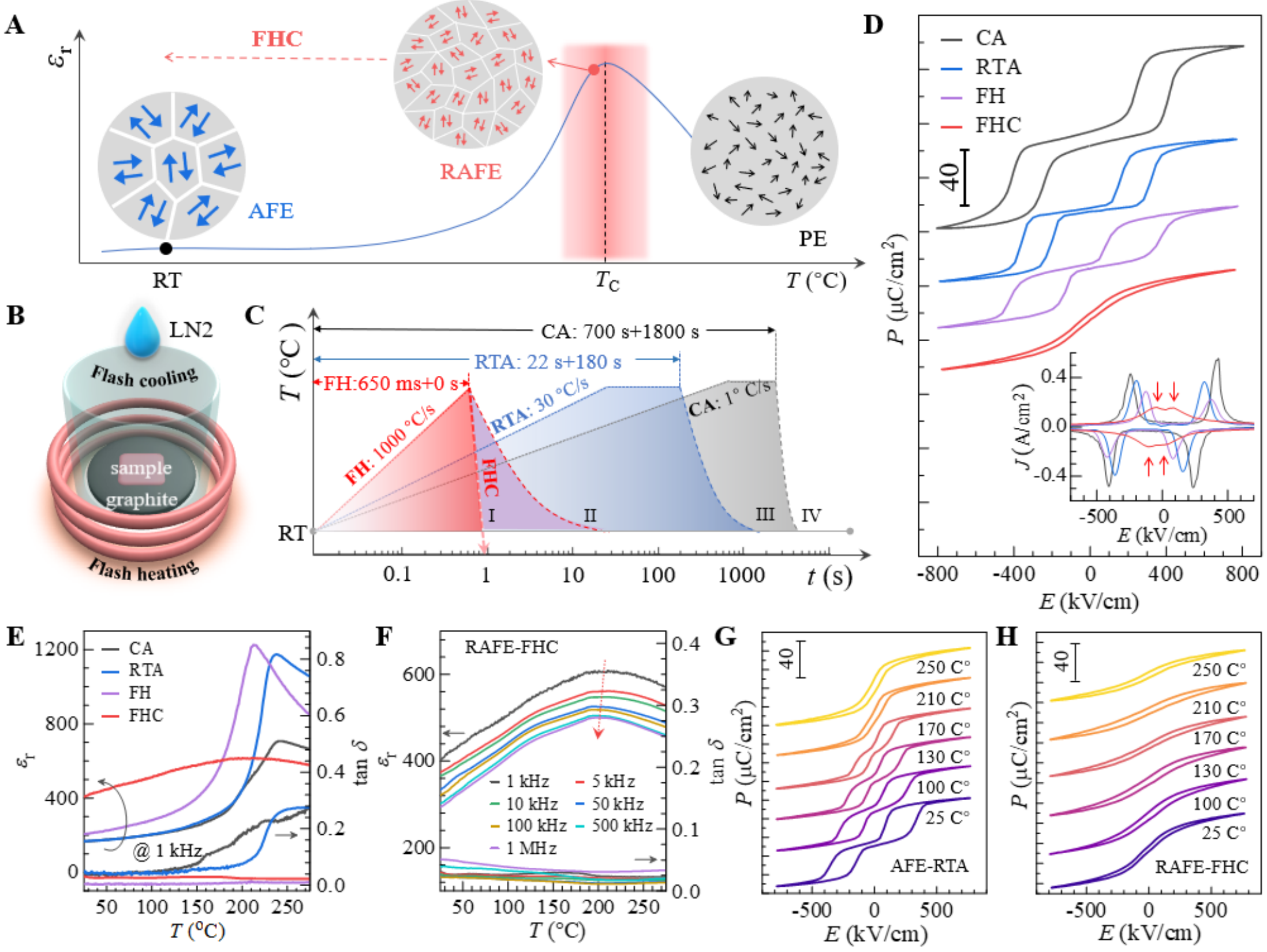

**Fig. 1. Fundamental design principles of relaxor antiferroelectric materials.** (**A**), The temperature-dependent dielectric permittivity ($\varepsilon_r$) of an antiferroelectric (AFE) material, depicted alongside representative schematic domain patterns at various temperatures. (**B**) Illustration of the experimental setup for flash heating and flash cooling (FHC), capable of achieving rates up to 1000 °C/s, enabling the synthesis of relaxor AFE materials within a single second. (**C**) A comparison of diverse heat treatment protocols, with treatment times spanning from less than 1 second for FHC to over 1000 seconds for CA. (**D**) *P–E* hysteresis loops for PZO films produced using different annealing techniques. The inset shows the corresponding switching current (*J–E*) curves. The red arrows indicate the peak-current positions for film processed by FHC. (**E**) Temperature dependent dielectric and loss for various films. (**F**) Temperature dependent dielectric for RAFE film at different frequencies. (**G** and **H**) *P-E* loops at various temperatures for AFE (**G**) and RAFE (**H**) films.

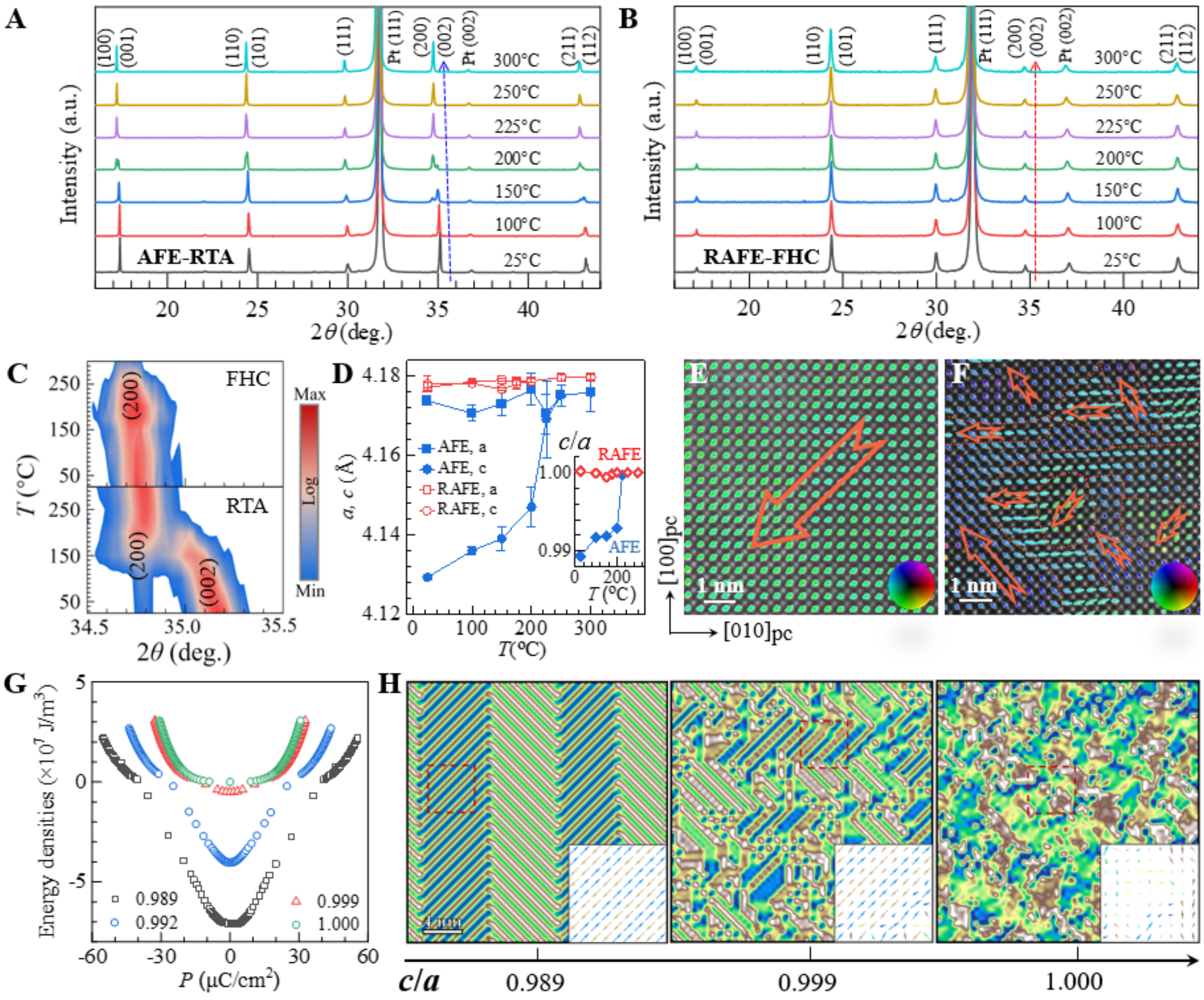

**Fig. 2. The crystal structural characteristics of AFE and RAFE films**. (**A, B**) XRD patterns at various temperatures for PZO films treated by RTA (**A**) and FHC (**B**). (**C**) The detailed view of (200)/(002) diffraction peaks. (**D**) Temperature dependent lattice constants. The inset represents the *c*/*a* ratio as a function of temperature. (**E** and **F**) HAADF-STEM images for AFE-RTA (**E**) and RAFE-FHC (**F**) films. The arrows at each ion show A-site cation displacement vectors in each unit cell. The orange arrows mark the domains of different orientations. (**G** and **H**) Free energy profiles (**G**) and domain patterns (**H**) with the increase of *c*/*a* ratio predicted by phase-field simulations. Insets in (H) show enlarged polarization vector fields from the red dashed boxes, offering a clearer view of local polarization distribution.

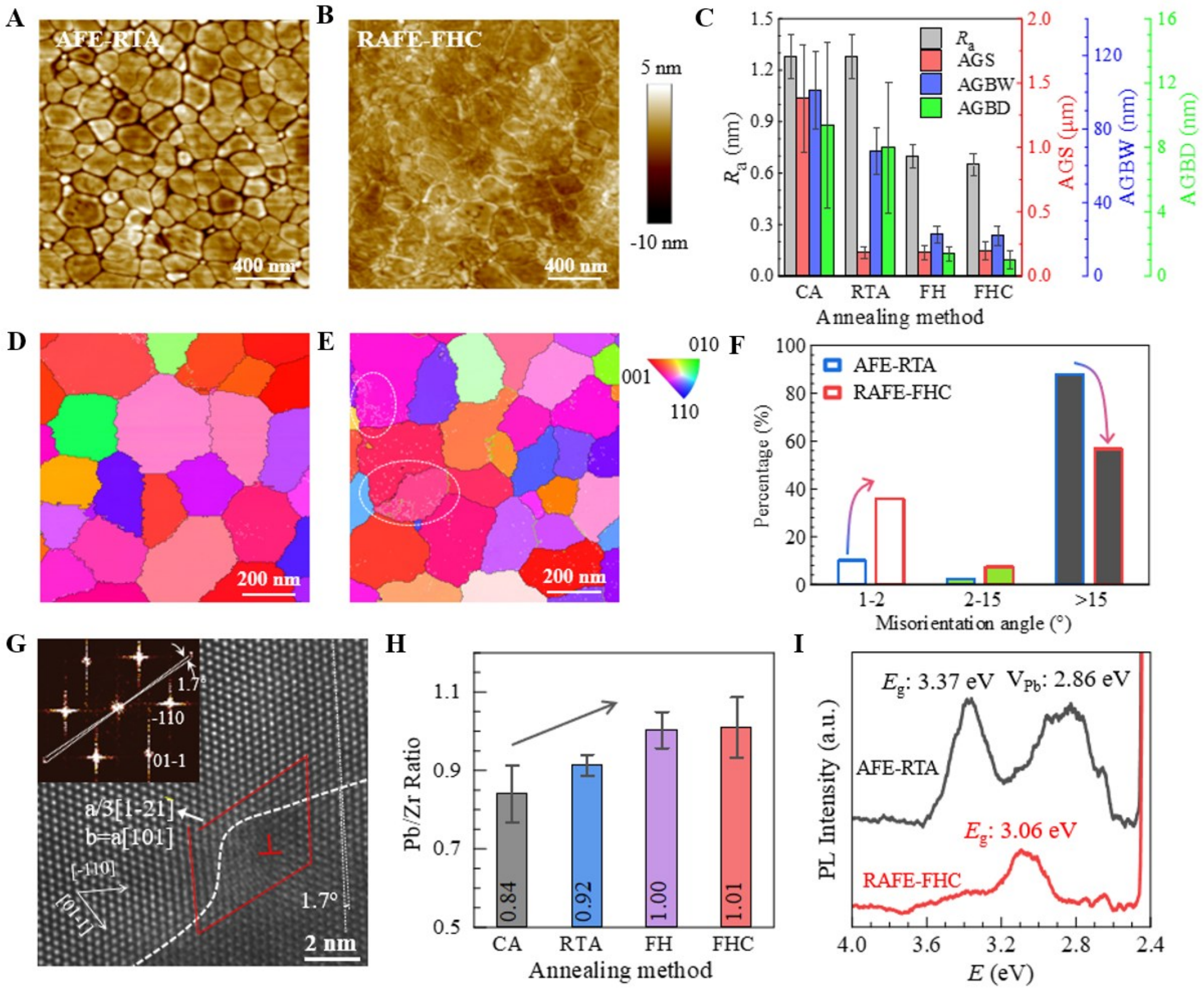

**Fig. 3. The microstructure characteristics.** (**A** and **B**) Topographic images for RTA-AFE (**A**) and FHC-RAFE (**B**) films. (**C**) Summary of surface roughness ($R_a$), average grain size (AGS), average grain boundary depth (AGBW) and width (AGBD) for various PZO films. (**D** and **E**) The IPF and GB situations for RTA-AFE (**D**) and FHC-RAFE films (**E**). (**F**) The misorientation angle statics derived from (**D**) & (**E**). (**G**) HAADF image of a sub-GB in RAFE film with a misorientation angle of 1.7°. Inset, the Fast Fourier transform diffraction patterns. (**H**) Pb/Zr atomic ratio statics derived from EDS analysis. (**I**) Photoluminescence (PL) spectrums of RTA-AFE and FHC-RAFE films.

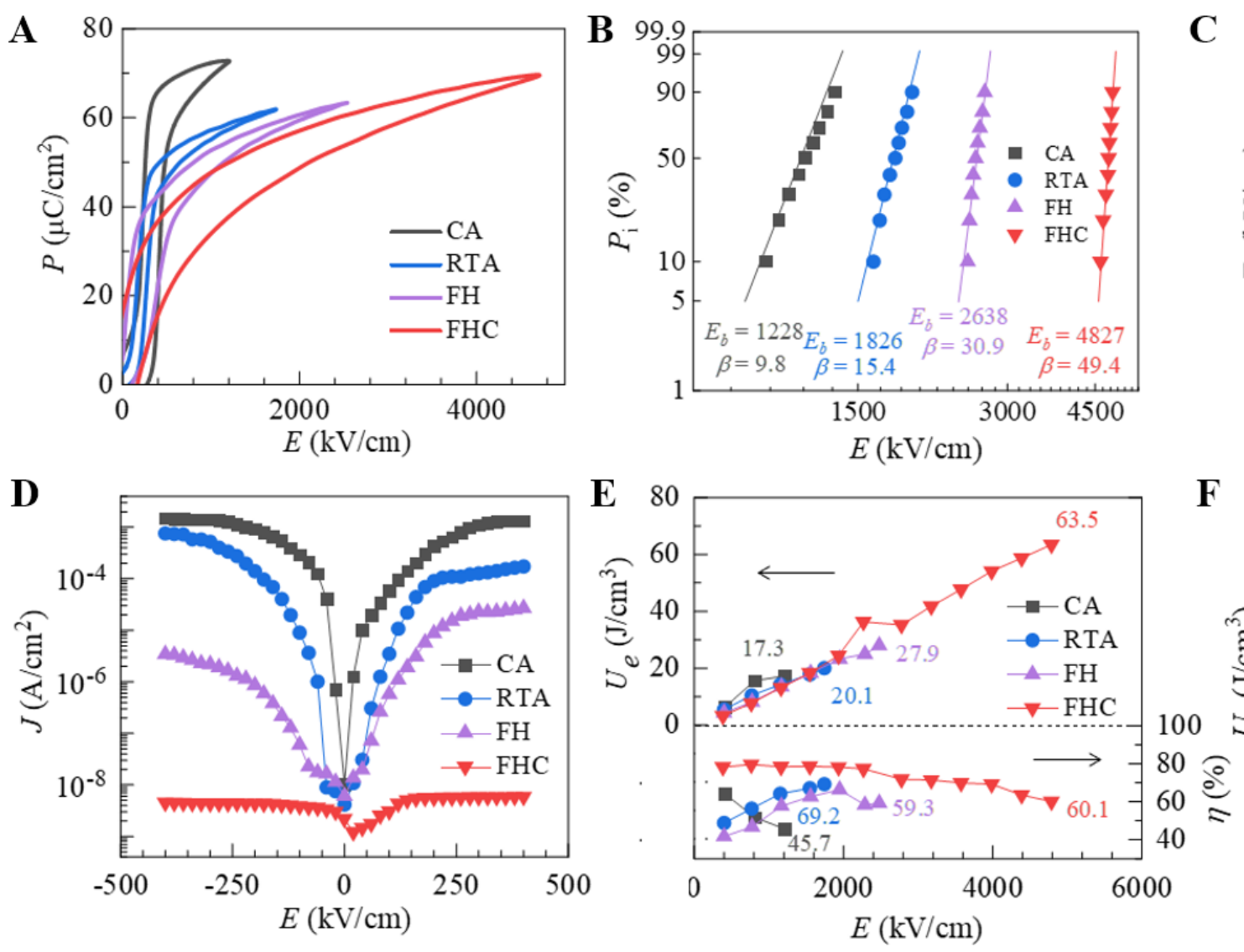

**Fig. 4**. **Polarization, electric, and energy storage properties.** (**A**) *P-E* hysteresis loops. (**B**) Two-parameter Weibull distribution analysis of the characteristic breakdown fields $E_b$. (**C**) $E_b$ *vs*. dielectric constant for typical dielectric materials. (**D** and **E**) Leakage current (**D**), and Energy density and efficiency (**E**) as a function of electric field for PZO films treated by different annealing methods. (**F**) Temperature dependent energy storage performance at 1500 kV/cm for AFE and RAFE films.

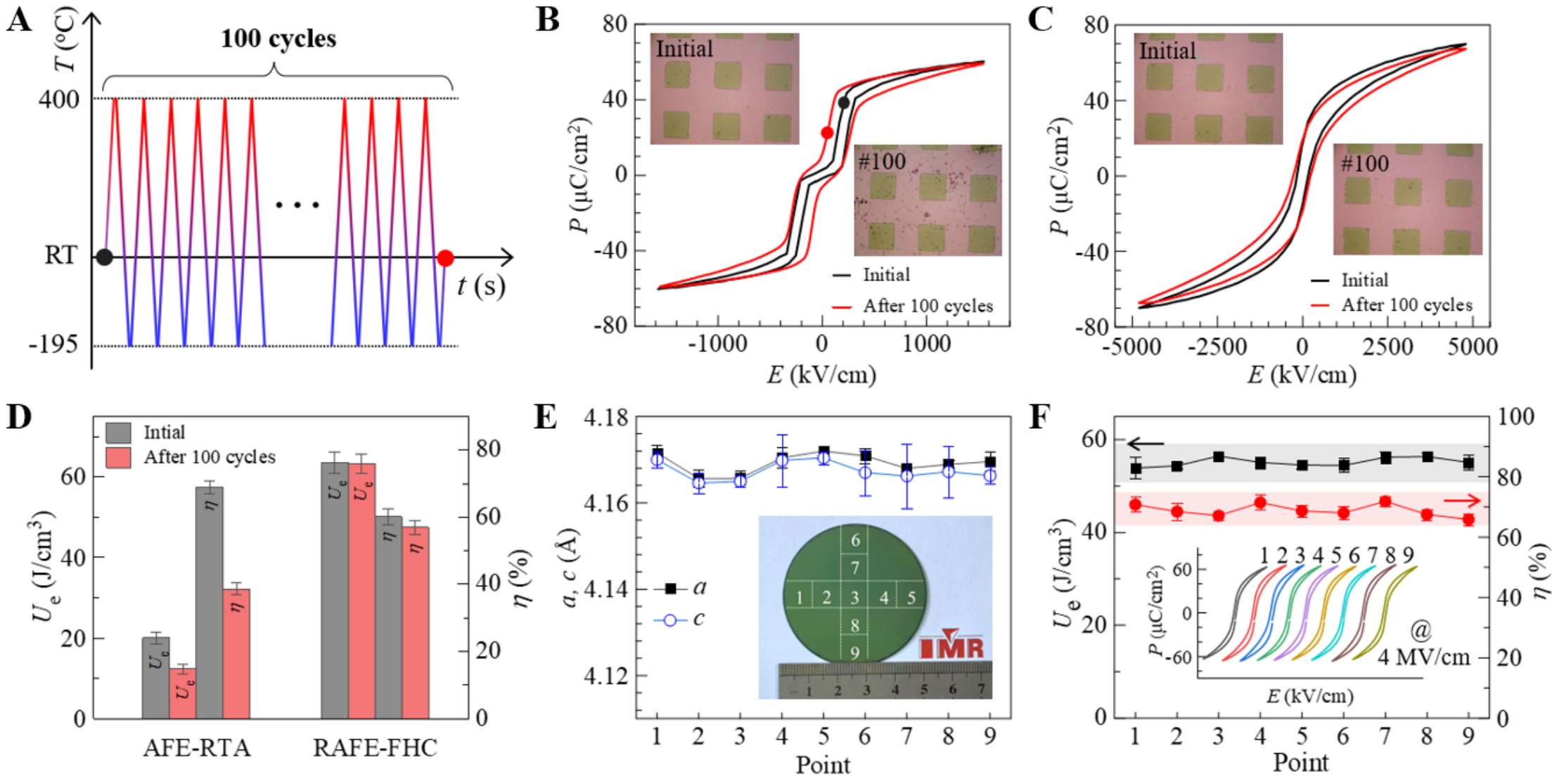

**Fig. 5**. **Thermal cycling reliability and scalability measurements.** (**A**) Protocol for thermal cycling test. (**B** and **C**) *P-E* loops of the pristine state and state after 100 thermal cycles for (b) RTA-AFE film and (c) FHC-RAFE film. Insets are the corresponding optical images of the films. (**D**) Energy storage performance in the pristine state and that after 100 thermal cycles. (**E**) Lattice constants on 9 typical positions across a 2-inch PZO film upon FHC treatment. Inset represents the optical image of the 2-inch film with the numbers indicating the locations of the measurements. (**F**) $U_e$ and $\eta$ on the 9 typical positions at 4000 kV/cm. Inset provides the corresponding *P-E* loops.

**Table 1. Energy storage performance of PZO films fabricated by various methods.**

| Method | State | Ann. time (s) | $P_m$ (μC/cm$^2$) | $E_b$ (kV/cm) | $U_e$ (J/cm$^3$) | $\eta$ (%) | Thermal stability ($\Delta U_e$, $\Delta\eta$%) | Ref. |
|---|---|---|---|---|---|---|---|---|
| **Magnetron sputtering** | AFE | 1800 | 45 | 700 | 12.5 | - | - | 31 |
| **PLD** | AFE | 3600 | 50 | 2000 | 46.0 | 63 | 24 to 250 ºC (≤47%, ≤38%) | 32 |
| **PLD & Ion implantation** | RFE | 3000 | 60 | 4493 | 62.3 | 60 | 25 to 125 ºC (≤5%, ≤5%) | 33 |
| **CSD & RTA** | AFE | 180 | 45 | 664 | 16.6 | 50 | 25 to 140 ºC (≤41%, ≤20%) | 30 |
| **CSD & Microwave radiation** | AFE | 180 | 56 | 765 | 14.8 | 59 | - | 34 |
| **CSD & FHC** | RAFE | 1 | 70 | 4827 | 63.5 | 60 | 25 to 250 ºC (≤3%, ≤3%) | This work |

# Supplementary Materials

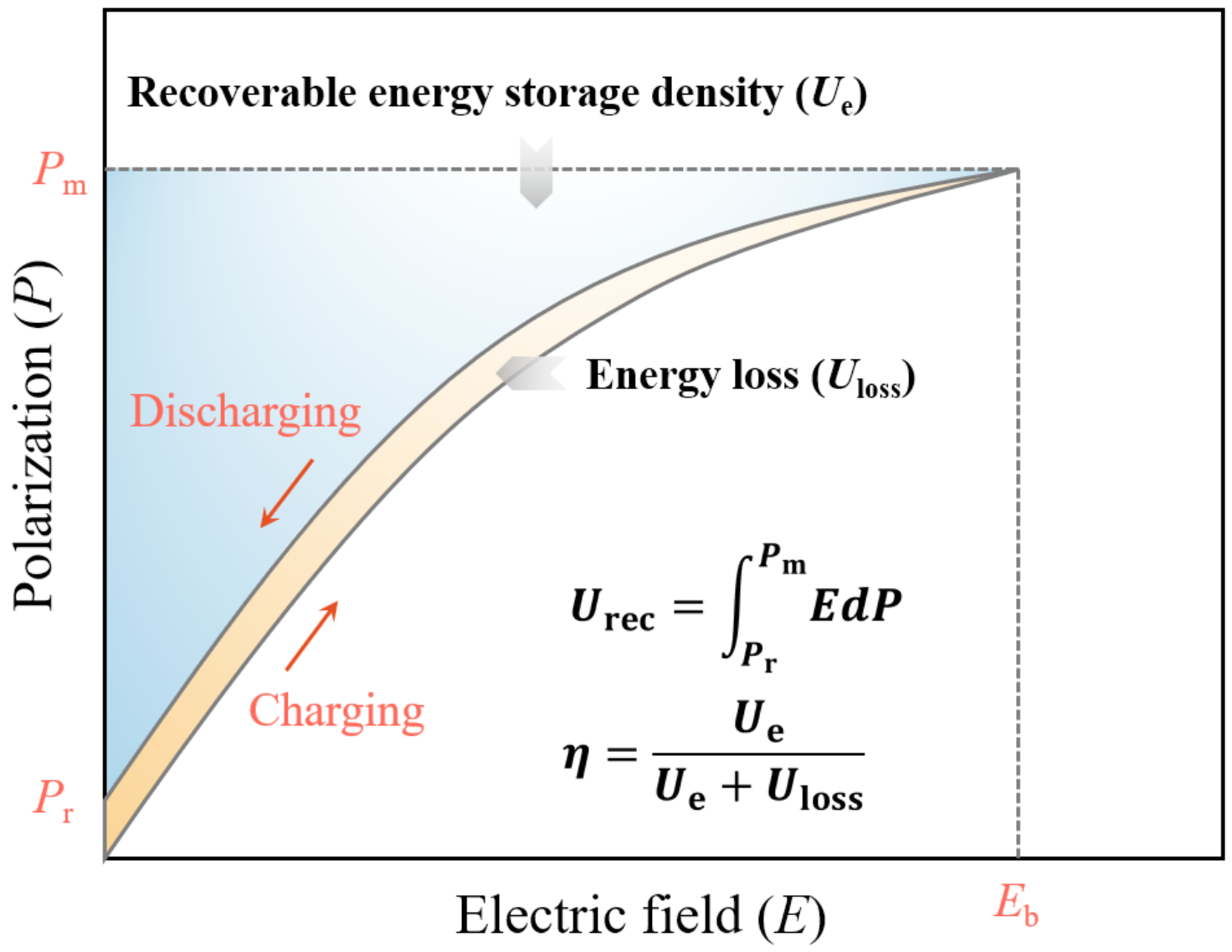

**Fig. S1. Working principle of energy storage capacitor.** The dischargeable energy density $U_e$ is depicted by the olive area, which is affected by both the polarization $P$ and the breakdown strength $E_b$ of the dielectric, as expressed by the equation where $P_m$ and $P_r$ indicate the maximum and remnant polarization, respectively. The pink area symbolizes the energy loss $U_{loss}$ from the hysteretic polarization switching during the charging/discharging cycle, while the energy storage efficiency $\eta$ is given by $U_e$ / ($U_e$ + $U_{loss}$).

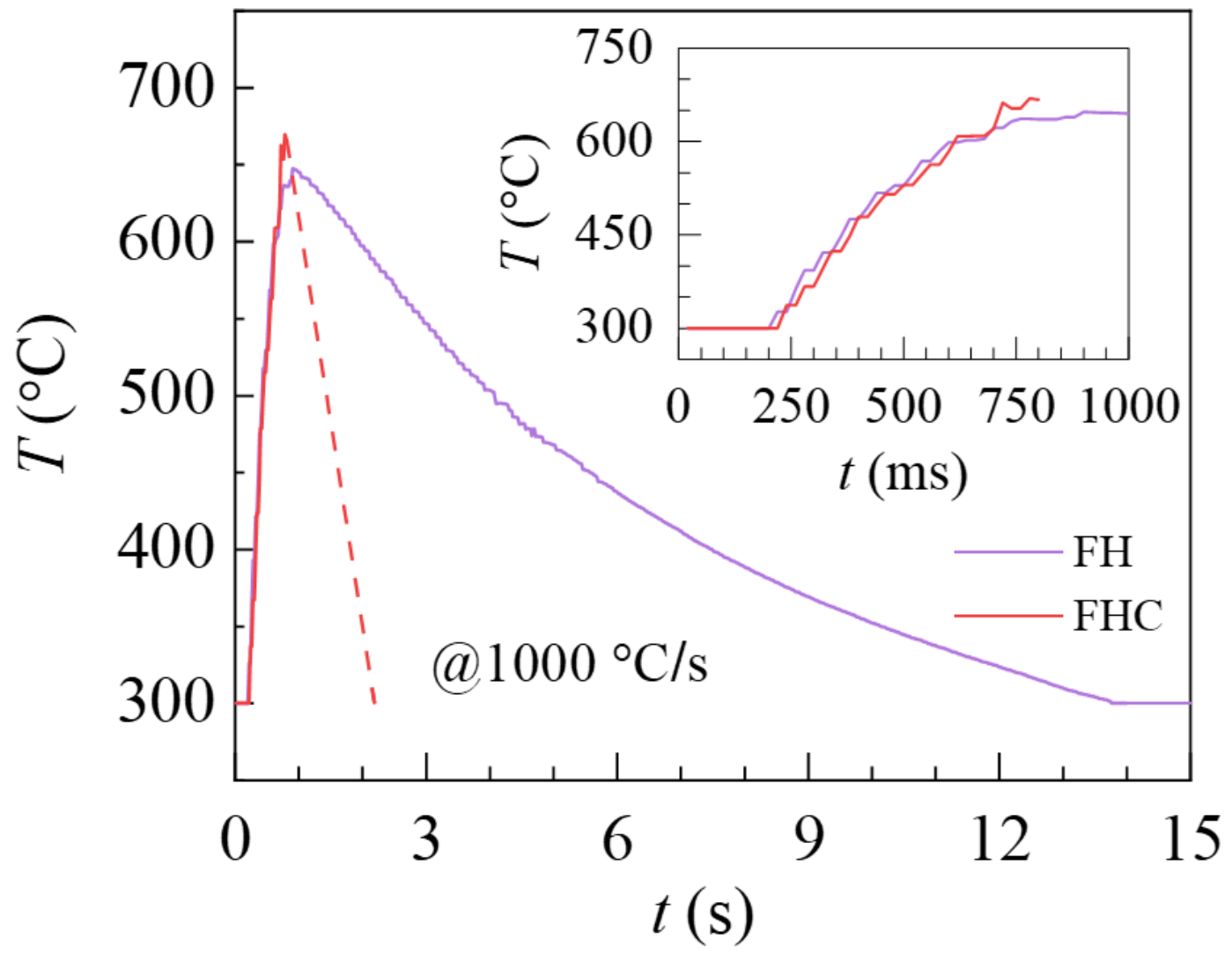

**Fig. S2. The sample temperature monitored by an infrared thermometer for FH and FHC process.** The inset provides a detailed view of the temperature rise during the heating process. Note 300 ºC is the lower detection limit of our infrared thermometer. Utilizing the Flash Heating and Cooling (FHC) and Flash Heating (FH) techniques, the heating rate can achieve an impressive 1000 ℃/s. It requires merely 650 milliseconds to reach the desired target temperature of 650 ℃. Moreover, by employing liquid nitrogen (LN2) quenching, FHC process achieves nearly instantaneous cooling within 1 second. While the FH treatment cool the sample in air with a cooling rate of ~30 ºC/s, requiring a cooling time of ~13 s.

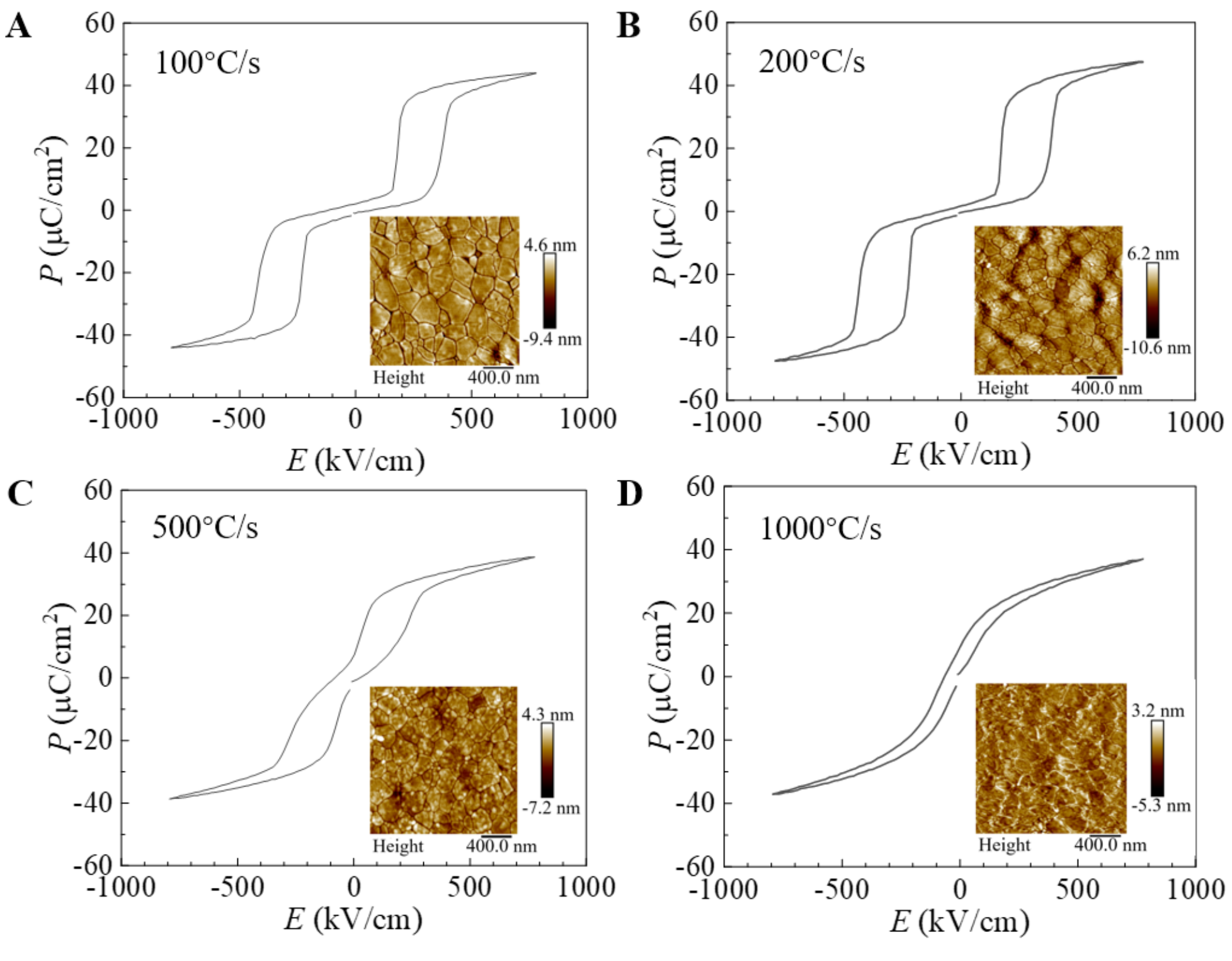

**Fig. S3. *P-E* loops with various heating rates used in FHC process.** (**A**) 100 °C/s, (**B**) 200 °C/s, (**C**) 500 °C/s and (**D**) 1000 °C/s. The insets present the AFM surface images of the corresponding films. Attaining a relaxor antiferroelectric state is possible via FHC treatment, however, this is conditional upon reaching an ultra-high heating rate of 1000 °C/s. Concurrently, the densification of grain boundaries experiences a marked improvement with increasing the heating rate.

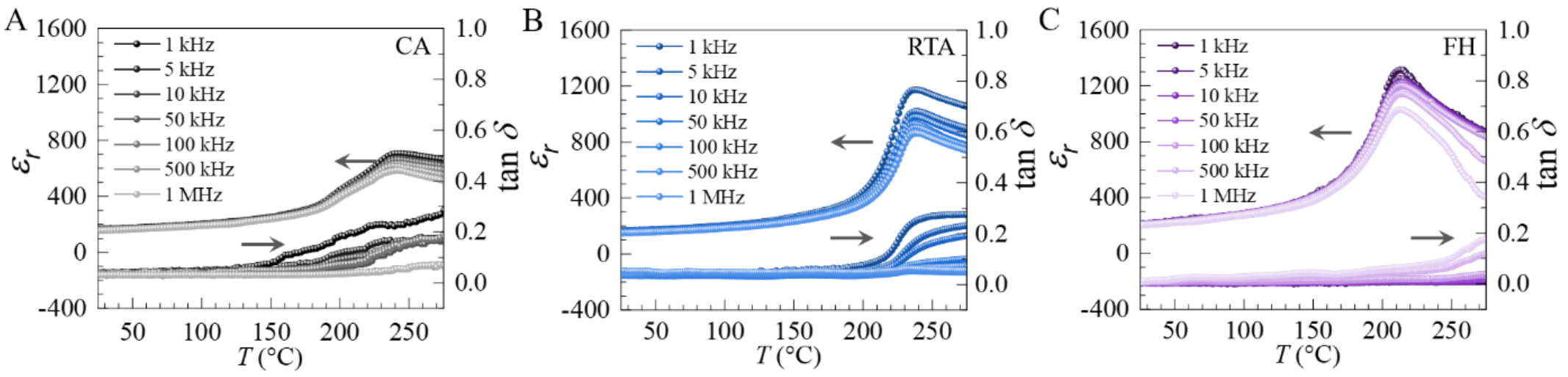

**Fig. S4. The temperature dependent dielectric spectra of PZO films treated by various processes.** A. CA. B. RTA. C. FH. Left, the dielectric constant ($\varepsilon_r$); and right, the dielectric loss (tan $\delta$). We applied an 1 V ac voltage during the measurement. The frequency ranges from 1 kHz to 1 MHz. With increasing the heating rate from CA, RTA to FH, distinct dialectic constant enhancement has been observed, particularly near the Curie temperature region.

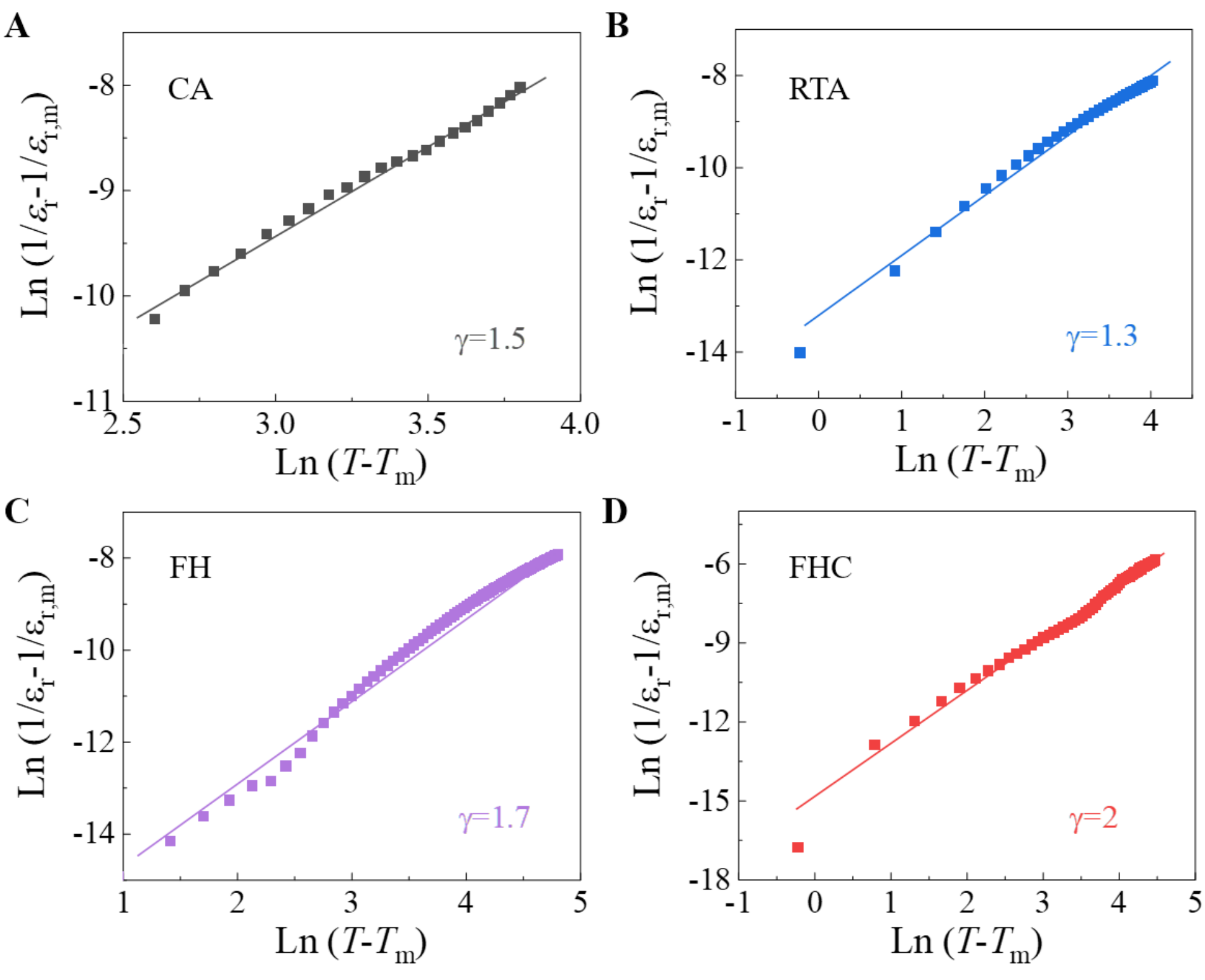

**Fig. S5. The relaxor diffuseness factor γ determined from temperature-dependent dielectric data at 1 kHz.** (**A**) CA, (**B**) RTA, (**C**) FH and (**D**) FHC. The solid lines are the fittings according to the modified Curie-Weiss law, $1/\varepsilon_r - 1/\varepsilon_{r,\ m} = (T - T_m)^\gamma/C$, where $\varepsilon_r$ is the permittivity at temperature $T$, $T_m$ is the temperature at which the permittivity reaches its maximum value $\varepsilon_{r,\ m}$, C is a constant, and γ is the relaxor diffuseness factor. γ ranges from 1 for normal antiferroelectric to 2 for an ideal relaxor antiferroelectric. This analytical approach provides a quantitative assessment of relaxor behavior, offering valuable insights for understanding the properties of PZO films.

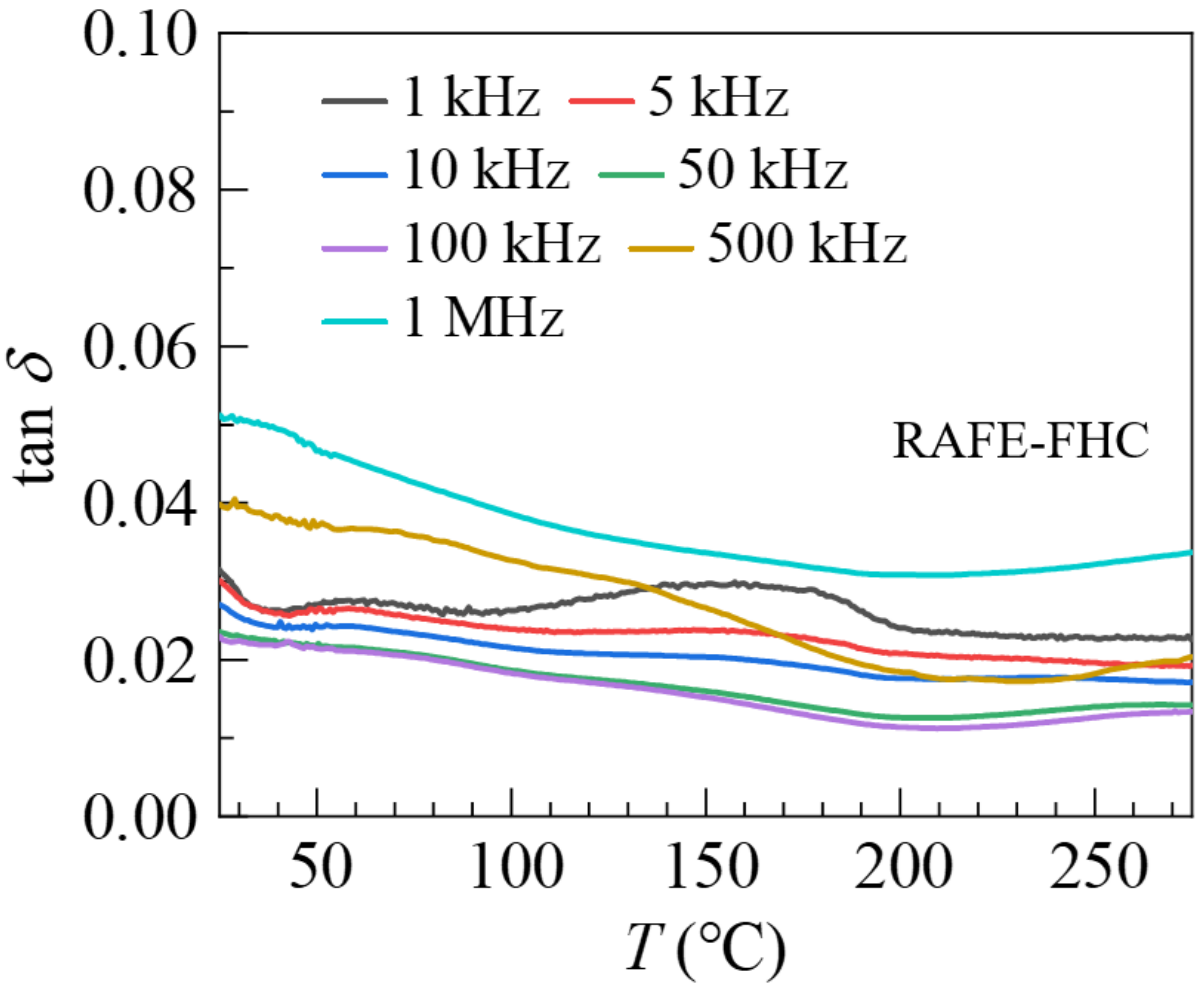

**Fig. S6. The enlarged dielectric loss as a function of temperature for PZO film treated by FHC process.** FHC-film shows minor frequency dispersion in dielectric loss. There are several reasons. First, abundant sub-grain boundaries and grain boundaries likely pin domain walls, impeding dipole rotation under ac electric fields. Second, the present FHC-film has a crystal structure closing to that of cubic paraelectric phase, suppressing the frequency dependence.

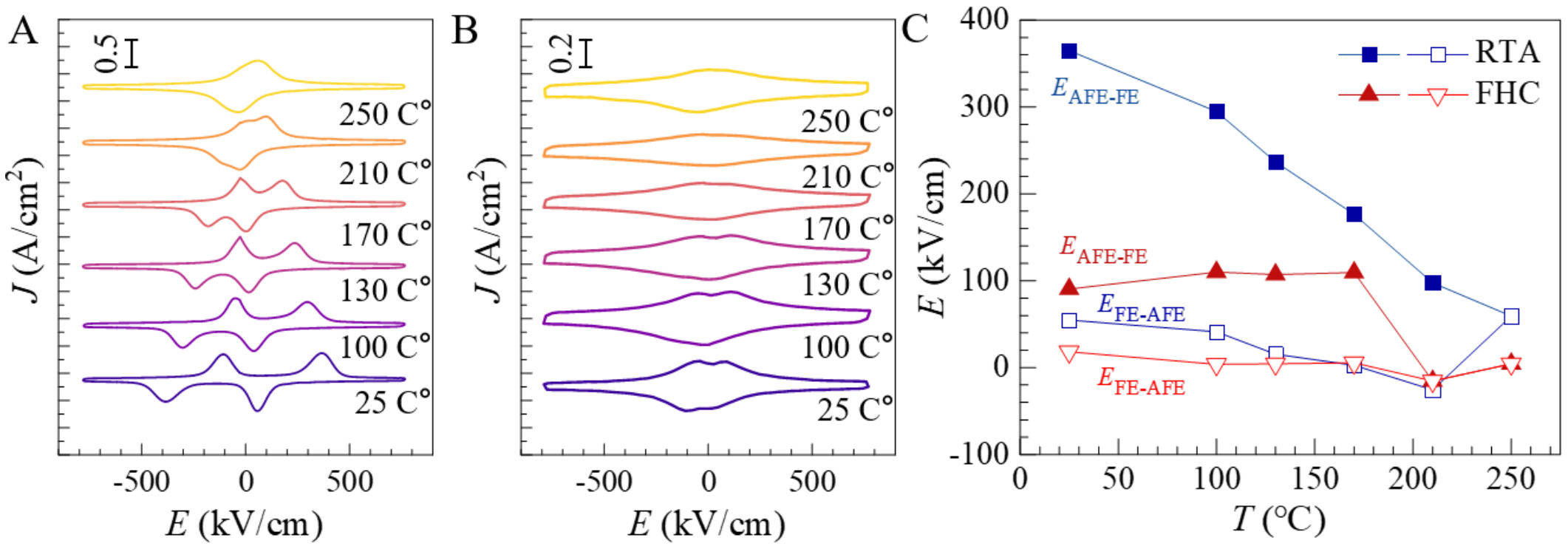

**Fig. S7. The temperature dependent AFE-FE transition behavior of PZO films**. Current switching curves at different temperatures for (**A**) RTA film, and (**B**) FHC film. (**C**) The AFE-FE transition field ($E_{\text{AFE-FE}}$) and FE-AFE back-transition field ($E_{\text{FE-AFE}}$) as a function of temperature for RTA and FHC film. The transition fields are determined from the peak positions of the switching current presented in **A** and **B**. Four distinct current peaks for RTA film at room temperature corelates to the sharp AFE-FE transition and FE-AFE transition in the double-hysteresis loop (**Fig. S7A**). These peaks gradually merge together with increasing the temperature above $T_c$ of ~ 238 °C. In contrast, though four current switching peaks exist for FHC film, their current densities are relative weak, and their peak positions are close to 0 kV/cm, due to the relaxer feature of FHC film (**Fig. S7B**). Therefore, FHC-film possesses smaller transition fields ($E_{\text{AFE-FE}}$ and $E_{\text{FE-AFE}}$) that are almost independent on the temperature, confirming the relaxor behavior of FHC-film and the similarity between RAFE and PE state.

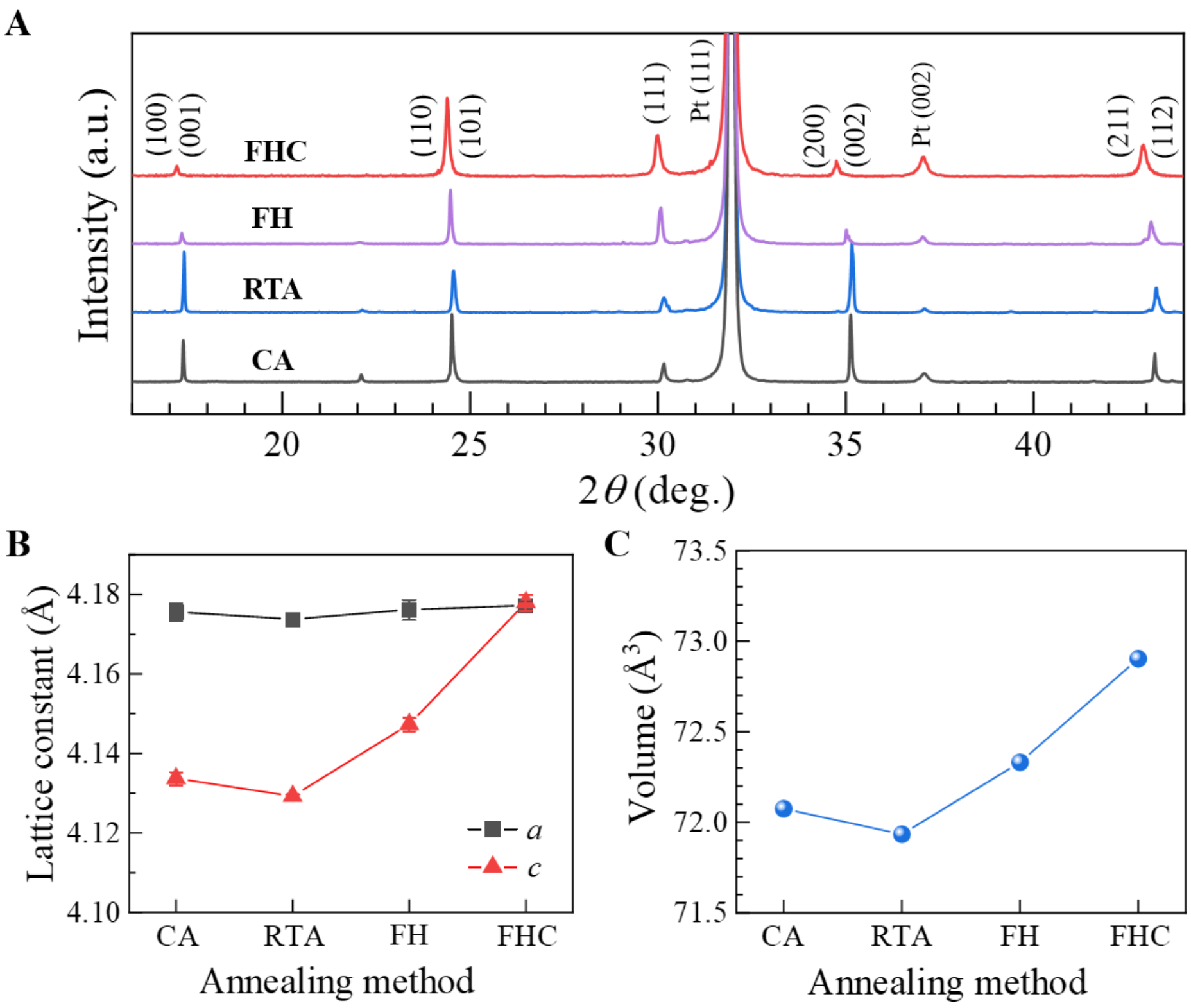

**Fig. S8. Synchrotron HRXRD analysis.** (**A**) HRXRD at room temperature by using an X-ray source with an energy of 10 keV. (**B** and **C**) The lattice constants (**B**) and the unit-cell volume (**C**) derived from (**A**). All films exhibit a polycrystalline nature with diffraction patterns arising from various crystal planes. As the heating rate is increased from conventional annealing (CA), rapid thermal annealing (RTA), to flash heating (FH) and flash heating and cooling (FHC), the diffraction peaks shift to lower angles. This shift corresponds to a volume expansion from ~ 72.1 $Å^3$ to 73 $Å^3$, representing an increase of 1.25%. This volume expansion is predominantly due to the expansion of the $c$ lattice parameter from ~ 4.135 Å to 4.180 Å, which is a 1.1% increase. The lattice constants $a$ and $c$ are nearly equivalent in the FHC film, suggesting a strong resemblance to the high-temperature paraelectric phase. This demonstrates that FHC process can effectively preserve the high-temperature crystal structure down to room temperature.

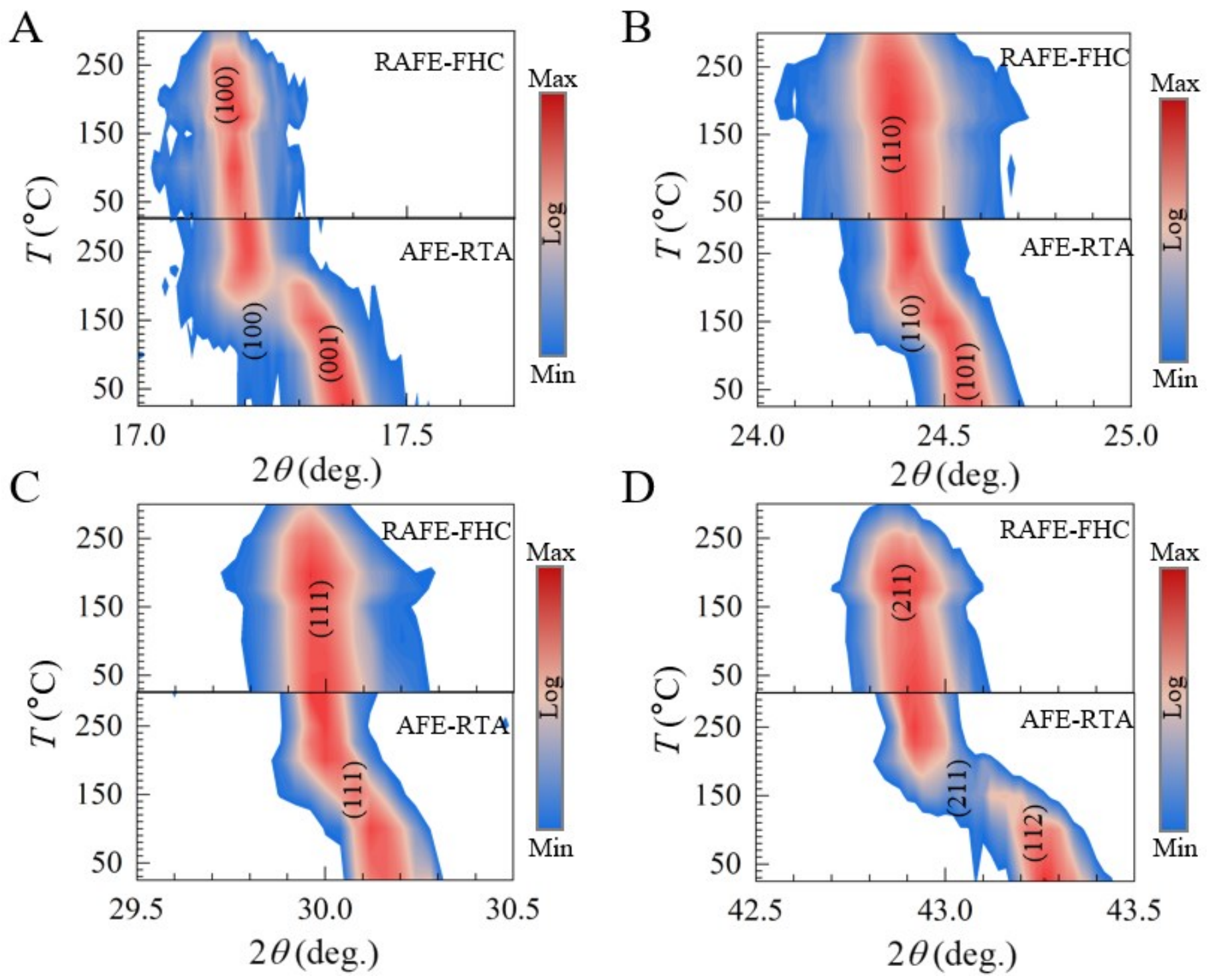

**Fig. S9. The temperature dependent diffraction peaks for various crystal planes.** A. (100)//(001). B. (110)//(101). C. (111) . D. (211)//(112). Distinct peak splitting is observed for (001) and (112) diffraction planes for RTA film, suggesting the tetragonal distortion because of the inequivalence of lattice *a* (or *b*) and *c*. The relative strong intensity of (001) over that of (100) indicate the preferred orientation exists in the film.

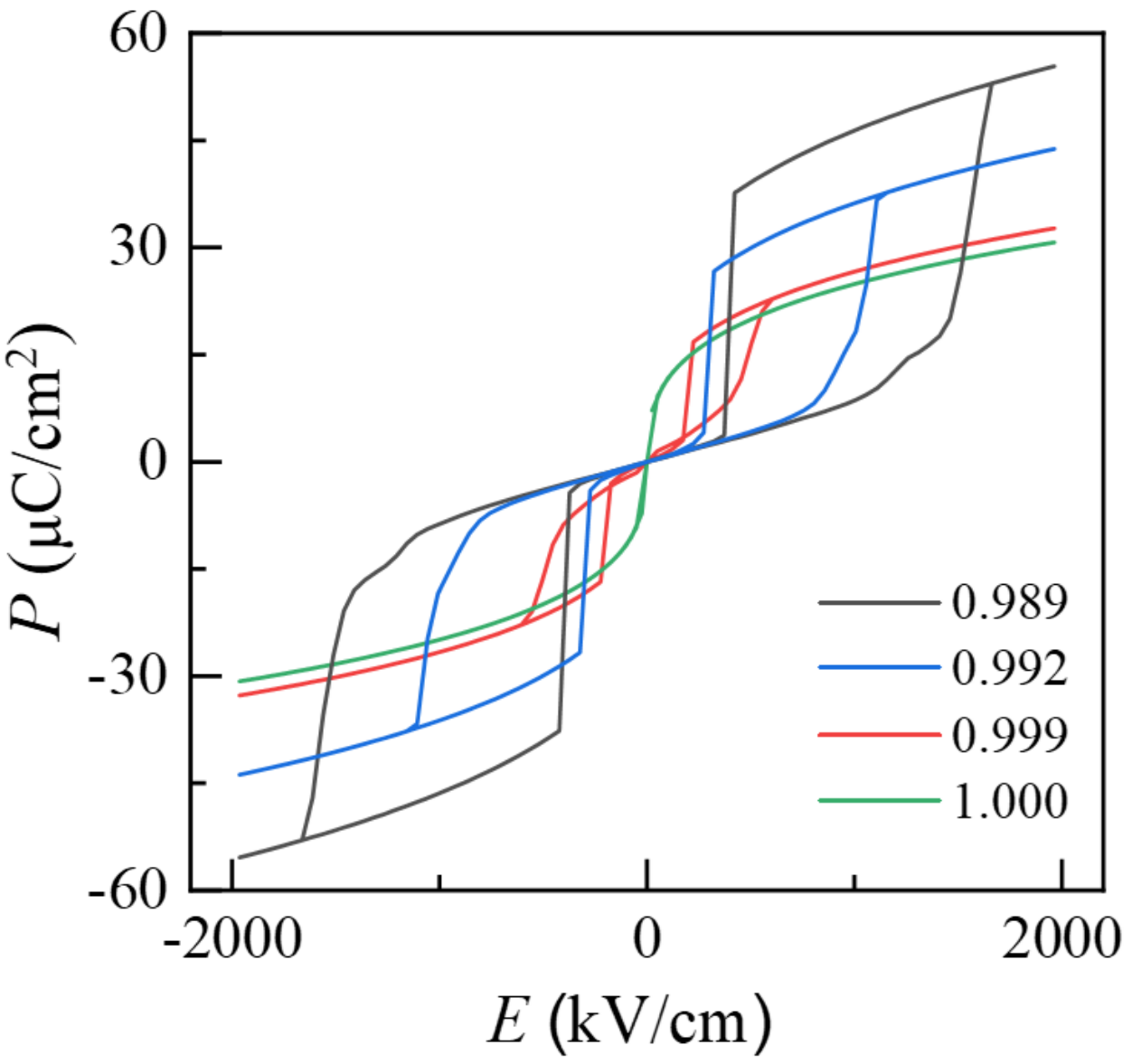

**Fig. S10. *P-E* loops at various c/a ratios predicted by phase-field simulation.** As the c/a ratio increases, the characteristic double-hysteresis *P-E* loops indicative of the antiferroelectric phase transits into the narrower, characteristic relaxor loops for the relaxor antiferroelectric phase.

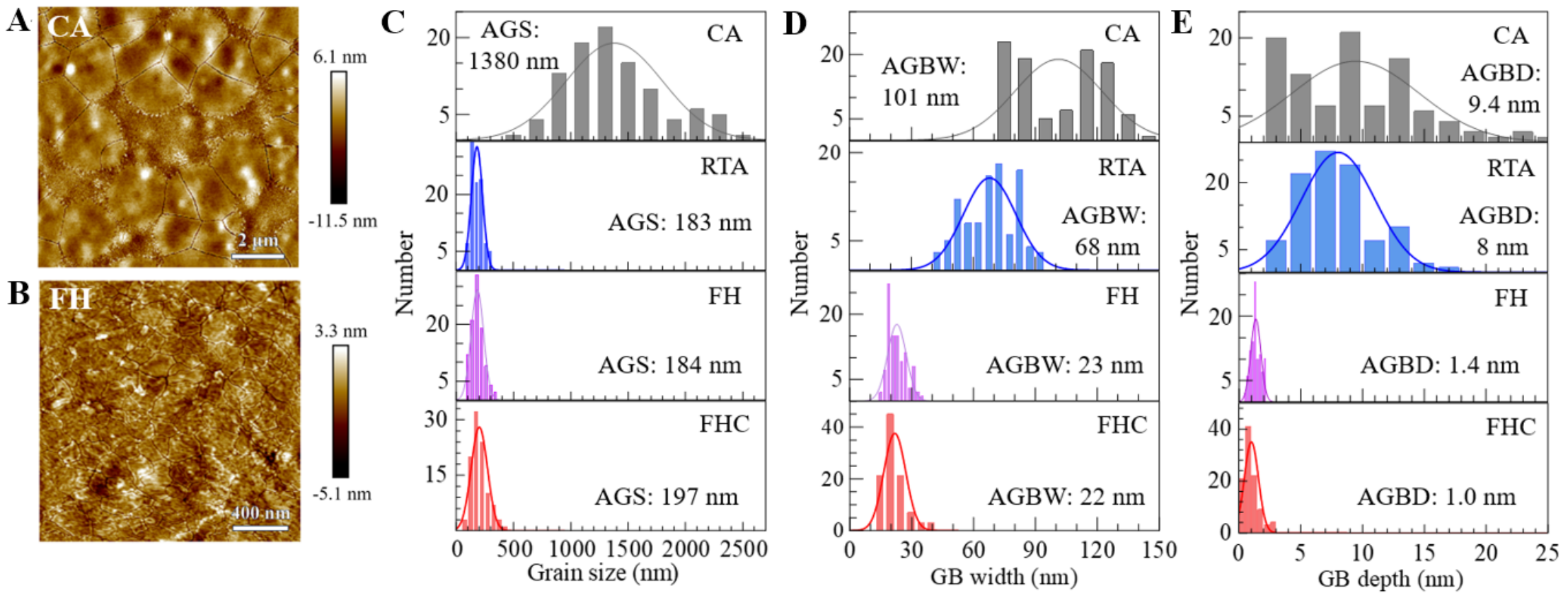

**Fig. S11. The grains and grain boundaries characteristics.** (**A** and **B**) The surface images of PZO films treated by CA (**A**) and FH (**B**). (**C** to **E**) Distribution of grain size (**C**), grain boundary width (**D**), and grain boundary depth (**E**) for PZO films treated by various processes.

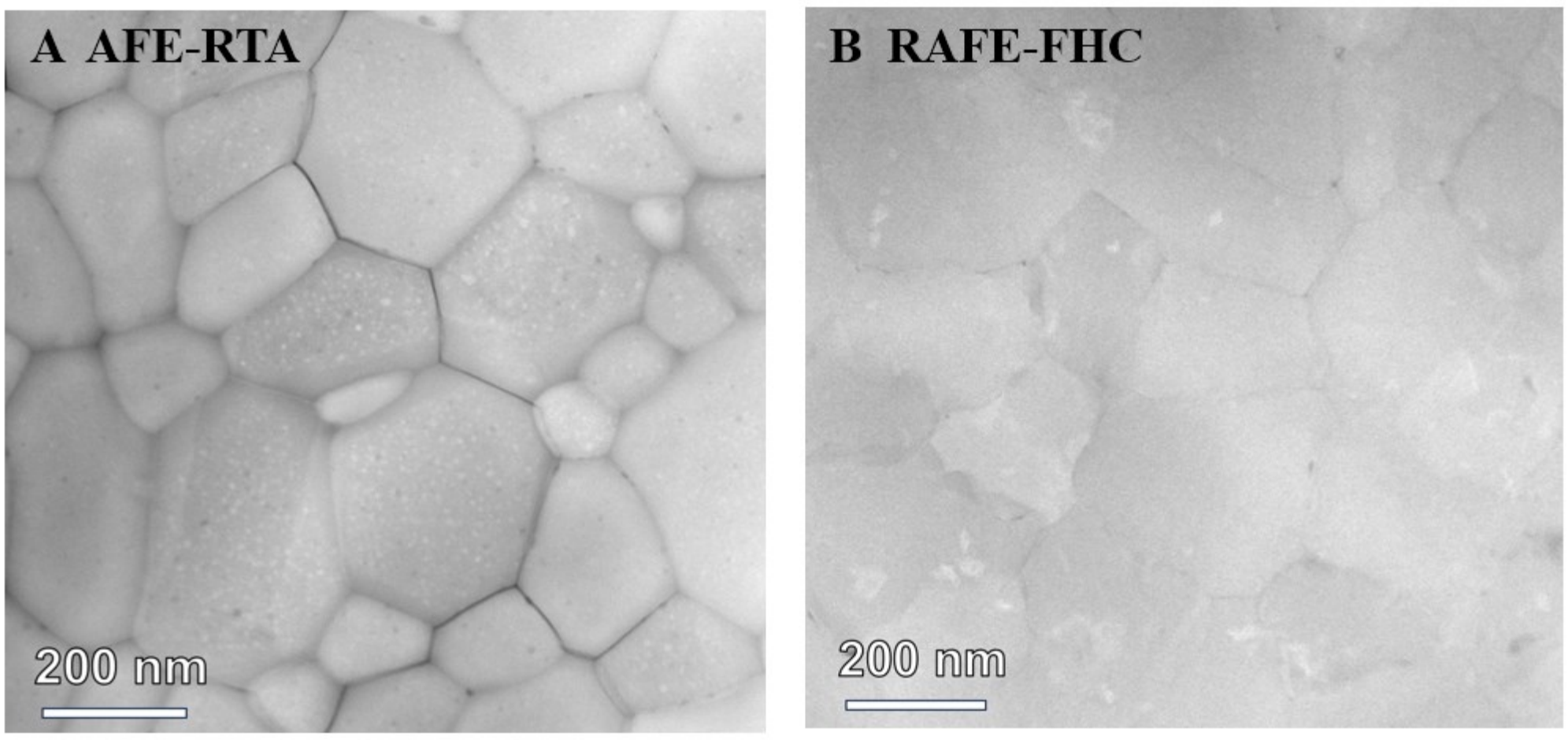

**Fig. S12. STEM morphology images of (A) AFE-RTA film and (B) RAFE-FHC film.** The AFE-RTA film exhibits polycrystalline characteristics, featuring regularly shaped and sharp-edged grains that are distinctly separated by grain boundaries. Conversely, the RAFE-FHC film, while also polycrystalline, presents a very smooth surface texture, with the grains not easily discernible due to the shallow and narrow grain boundaries. This results in an exceptionally compact and uniform surface.

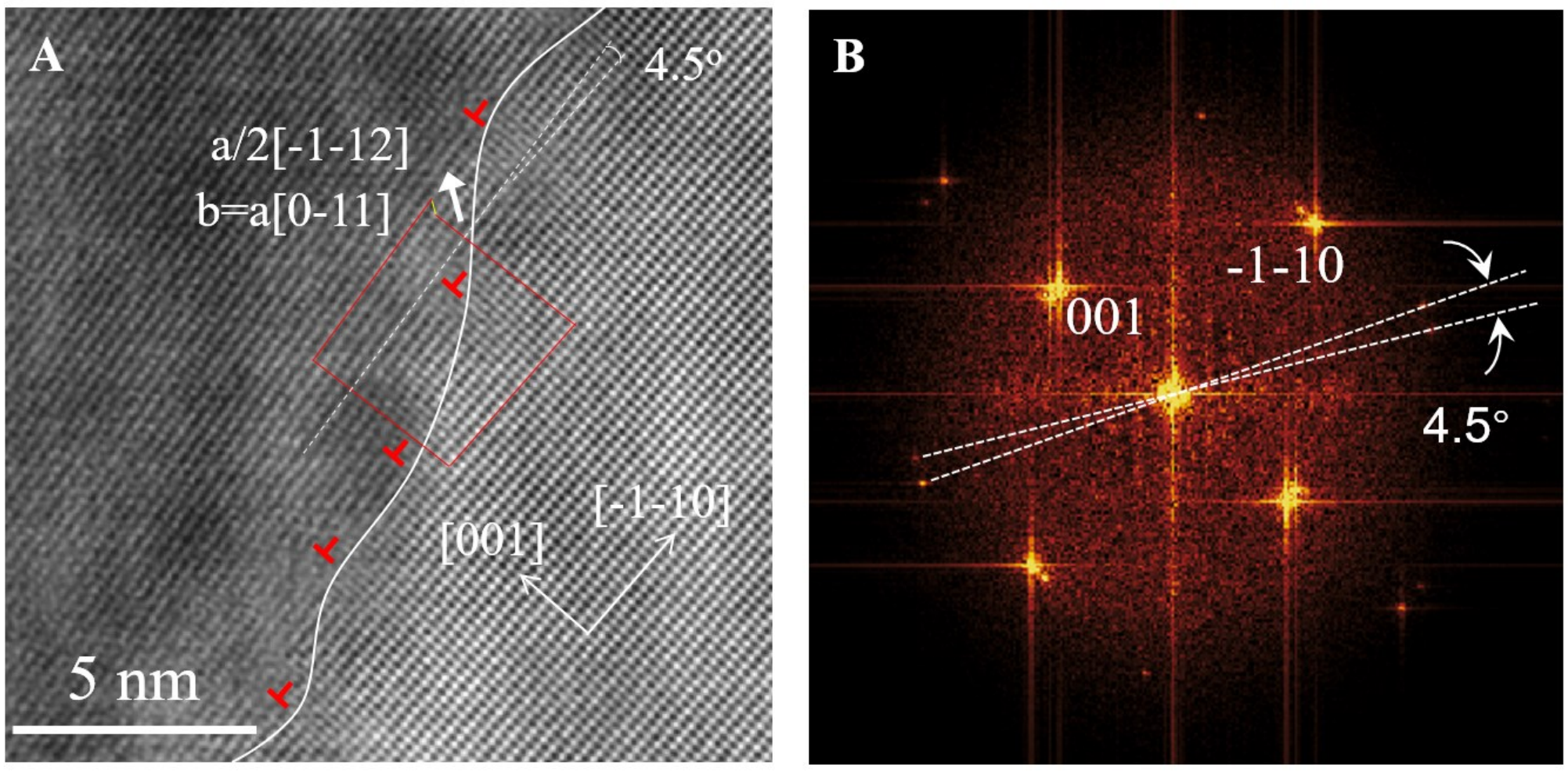

**Fig. S13. HRTEM image of a low-angle GB in RAFE-FHC film.** (**A**) The HAADF-STEM image. The solid line marks the GB position. (**B**) The diffraction pattern derived from (**A**) via Fast Fourier transformation. Two distinct sets of diffraction spots are evident in the fast FFT pattern, enabling us to determine the misorientation angle of this GB to be 4.5°. Following the grain boundary as indicated by the white line in (**A**), we have identified a series of perfect dislocations with a Burgers vector of a [0-11]. This finding indicates that the formation of GB is a consequence of dislocation pile-up at the boundary.

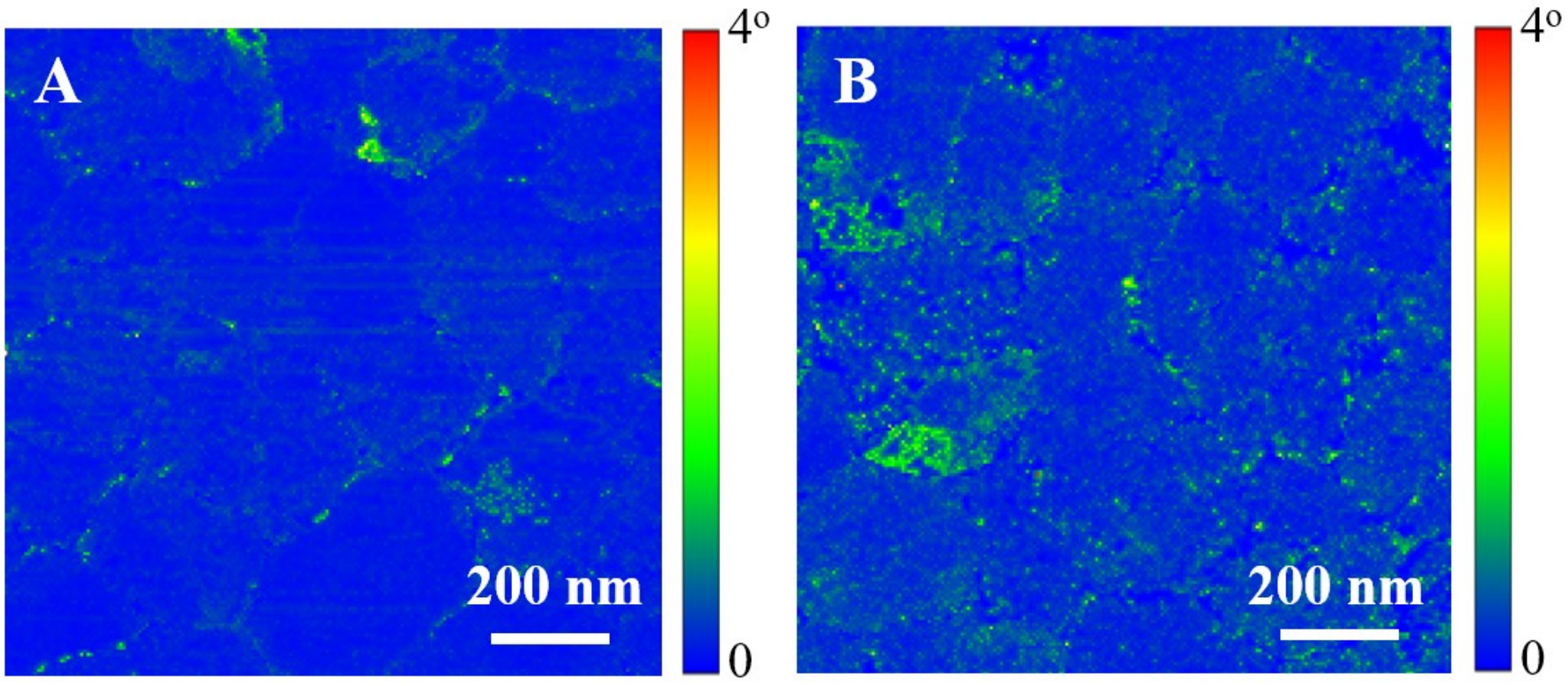

**Fig. S14. Kernel Average Misorientation (KAM) mappings for PZO films treated by (A) RTA process, and (B) FHC process.** Compared to RTA-film, larger intra-grain misorientation angles appear in FHC-film, particularly at grain boundaries and within sub-grains, suggesting higher residual stress levels in FHC-film.

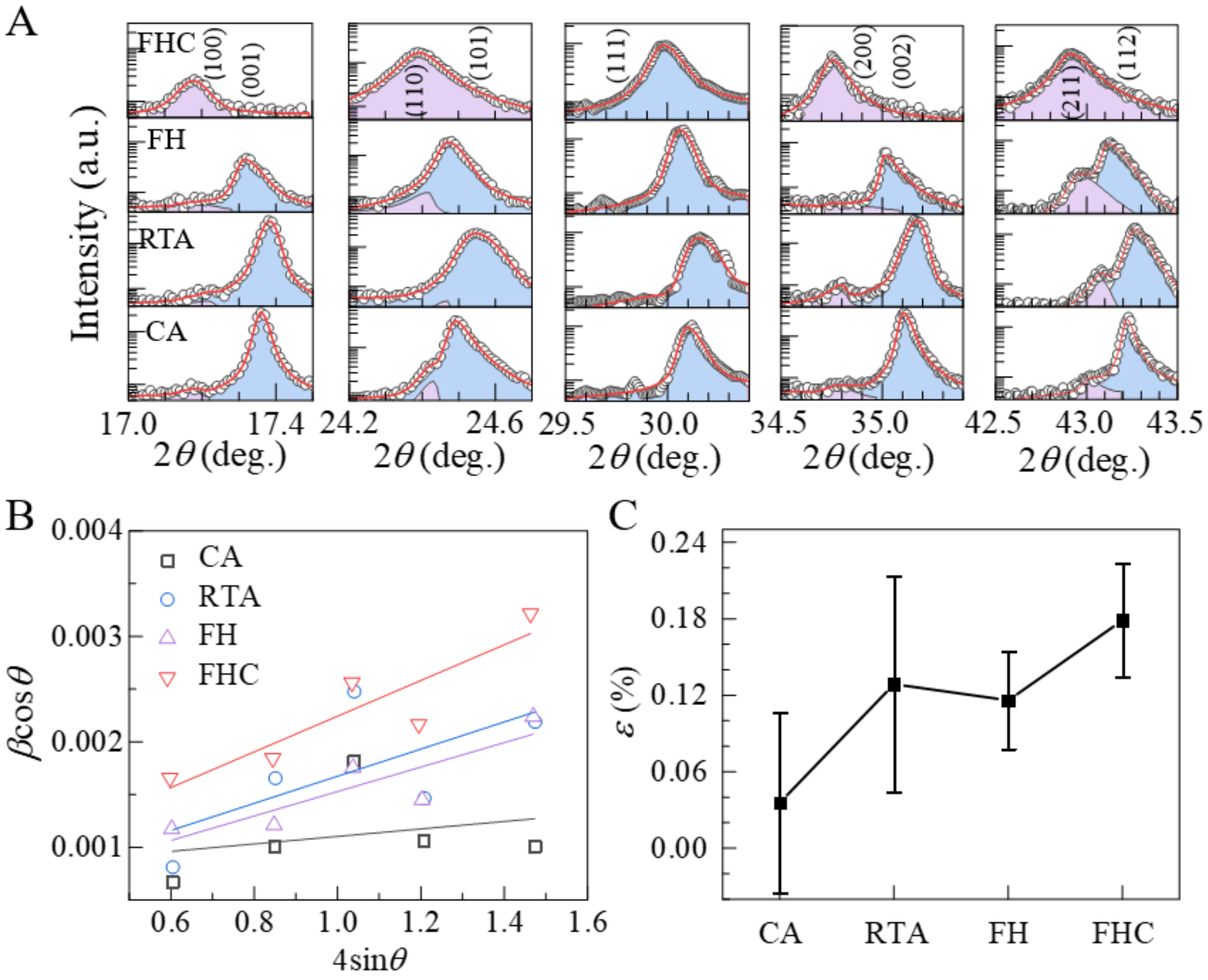

**Fig. S15. X-ray diffraction broadening analysis.** A. The enlarged XRD peaks at (100)/(001), (110)/(101), (111), (200)/(002), and (211)/(112) for PZO films treated by CA, RTA, FH, and FHC processes. B. Williamson-Hall plot ($\beta\cos\theta$ *vs.* $4\sin\theta$) for all samples with $\beta$ derived from A. C. Microscopic strain $\epsilon$ derived from the slope of the curves shown in B.

To quantify the internal stress, we analyzed the peak broadening (characterized by Full Width at Half Maximum, FWHM) of room-temperature XRD patterns across different films (**Fig. S8A**). This is based on the fact that strain-induced lattice distortions would widen the diffraction peaks. However, we mention that this broadening could also arise from finite crystallite size (Scherrer effect) and instrumental broadening. Here we use Williamson-Hall (W-H) method to decouple these contributions, which states that the total broadening of the XRD peak follows equation $\beta\cos\theta = \frac{K\lambda}{D} + 4\varepsilon\sin\theta$, with the first term represents the peak broadening induced by the crystallite size and the second term by the strain. $\beta$ is the FWHM of the diffraction peak, $\theta$ is the diffraction peak, $K$ is the Scherrer factor, $\lambda$ is the wavelength, and $\epsilon$ is the strain. As evidenced in **Fig. S15A**, the enlarged diffraction peaks of PZO films [(100)/(001), (110)/(101), (200)/(002), and (211)/(112)] reveal distinct peak splitting for CA-, RTA-, and FH-processed films-a characteristic signature of tetragonal distortion. Whereas, FHC film shows minor peak splitting. By fitting these diffraction peaks with Pseudo-Voigt function, we obtain $\beta$ values and plot $\beta\cos\theta$ *vs.* $4\sin\theta$ as shown in **Fig. S15B**, from which we derive the microscopic strain $\epsilon$ from the slope of each curve and summarize them in **Fig. S15C**. $\epsilon$ increases from 0.036% for CA, 0.13% for RTA, 0.12% for FH, and ultimately to 0.18% for FHC, quantitatively demonstrating the critical role of heating/cooling rate in tailoring the microstructure of PZO films. As evidenced by our analysis, the flash heating/cooling during FHC process generates a large amount of sub-grain boundaries (**Fig. 3F**) together with dislocation networks (**Fig. 3G & Fig. S13**), resulting in more statistical fluctuations in lattice constants, manifesting as enhanced microscopic stress that stabilize the high-temperature cubic-like phase.

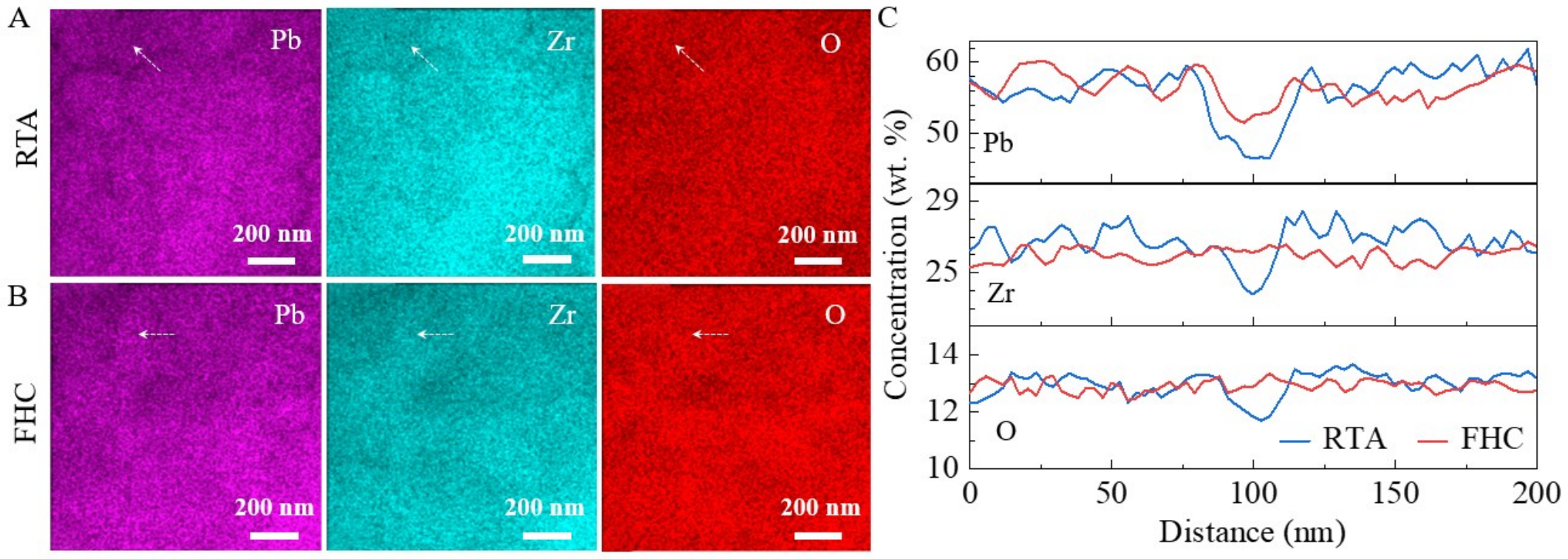

**Fig. S16. EDS mapping of PZO films.** (A) RTA film and (B) FHC film. (C) The element line scans along the typical grain boundaries (indicated by dashed arrows in A and B) for RTA and FHC films. While both films show homogeneous element distributions within the grains, FHC-film shows minor Pb reduction on the grain boundaries as indicated by the minor intensity fluctuations across the grain boundaries, compared with that of RTA-film. This has also been confirmed from the line scan across the grain boundaries, as displayed in **Fig. S16C**.

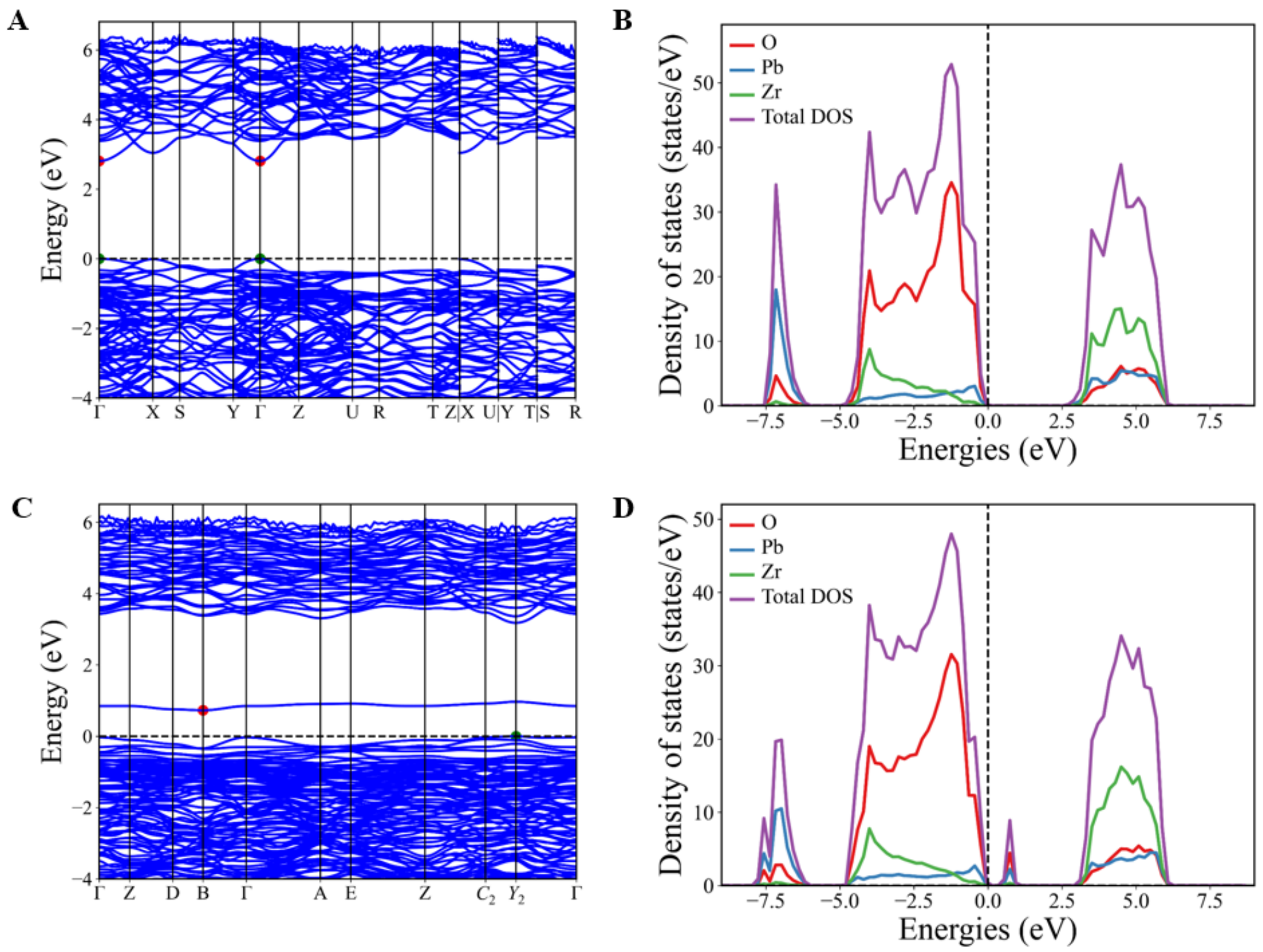

**Fig. S17. Band structures and defects.** (**A**) Band structure and (**B**) Partial density of states (DOS) of PZO film without considering point defects. (**C**) Band structure and (**D**) Partial DOS of PZO with Pb vacancies.

The calculated band structure and DOS in (**C**) and (**D**) confirms that Pb vacancy introduces an acceptor level at ~0.7 eV above the valence band maximum (VBM). The 0.2 eV discrepancy between this theoretical value and experimental observations can be attributed to the limitations of the PBE functional method, which does not adequately account for strong-correlation electronic systems. The observed O orbital contribution in the in-gap states is attributed to local structural and electronic perturbations around the Pb vacancy, and such coupling between Pb and O is consistent with defect configurations reported in perovskite systems, where Pb-O defect complexes are known to form under Pb-deficient conditions (*47, 48*).

The bandgap of the RAFE film is approximately 3.06 eV, which is ~0.3 eV smaller than that of the normal AFE-PZO film. This reduction is ascribed to the combined effect of lattice constants and Pb vacancy concentrations, as supported by the bandgap values summarized in **Table S2**.

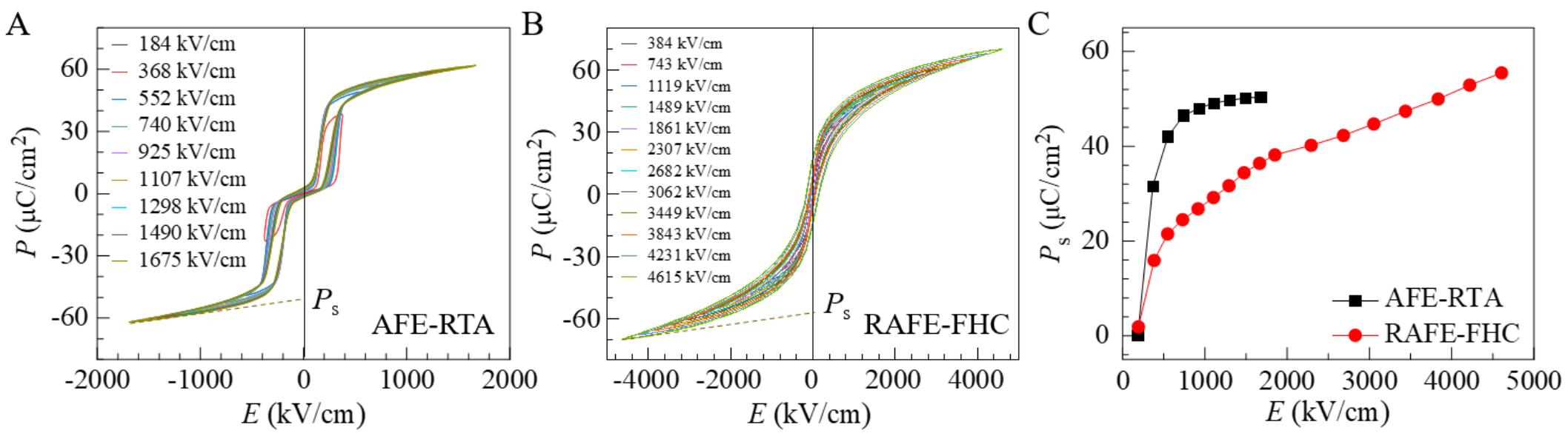

**Fig. S18. *P-E* loops under various maximum electric fields.** A. for AFE-RTA film, B. for RAFE-FHC film, and C. the saturation polarization $P_s$ as a function of electric field. Here $P_s$ is determined by extrapolating the high-field linear part to zero field as depicted by dashed lines in A and B. For small electric field of ~200 kV/cm, $P_s$ equals to 0 both for RTA and FHC films because they are in the AFE or RAFE ground state. Increasing the electric field will see the sharp increase of $P_s$ for RTA film because of the sharp field-induced AFE to FE transition. In contrast, the increase of $P_s$ in FHC film is slowly because of its relaxor feature. However, $P_s$ of FHC film will finally exceeds that of RTA film when increasing the electric field further, because it can sustain larger electric field without breaking.

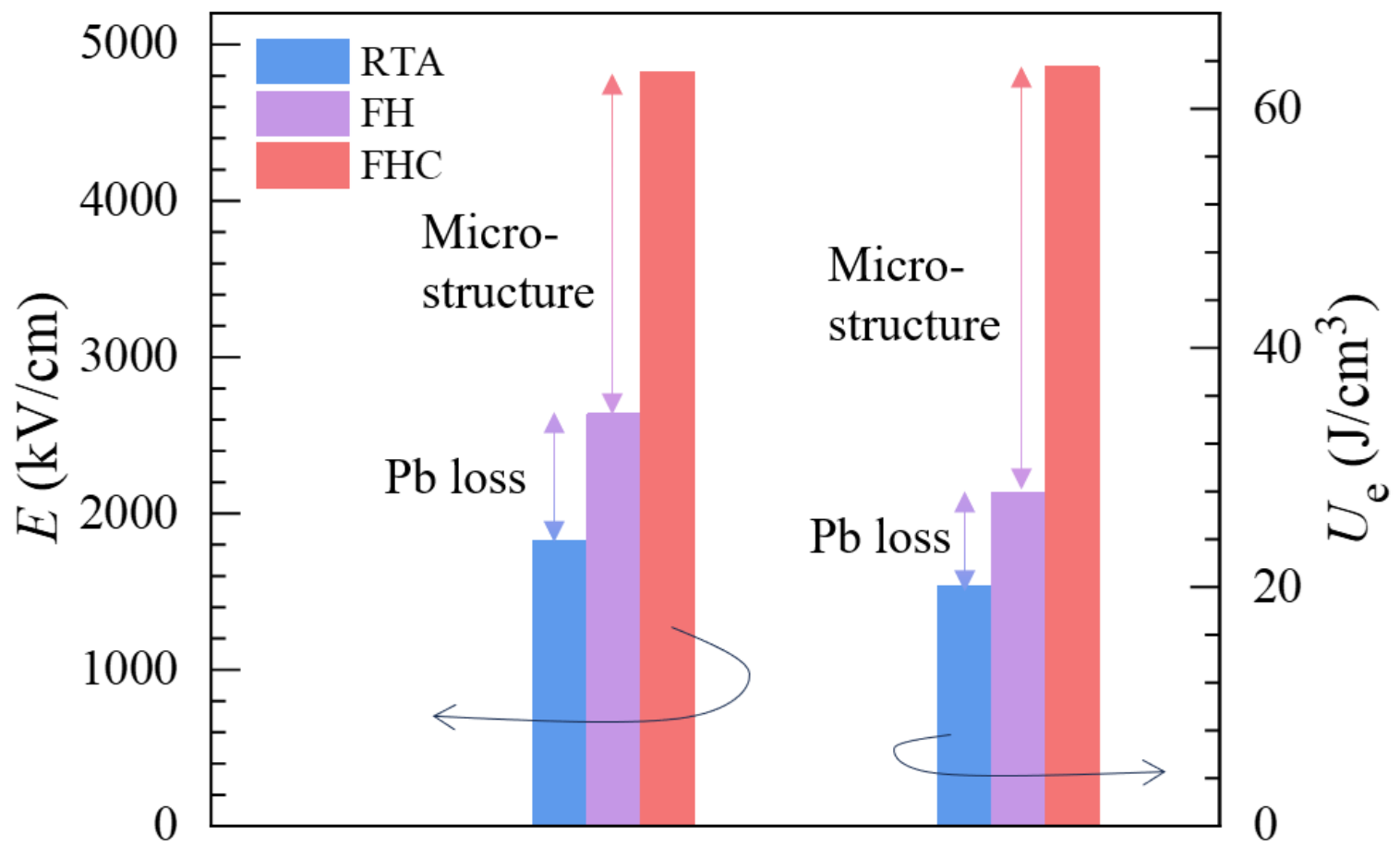

**Fig. S19. Role of Pb deficiency vs. microstructure in energy storage performance.** Breakdown strength ($E_b$, left) and energy storage density ($U_e$, right). Microstructural factors include sub-grains and nanodomains. **Fig. 3H** shows the Pb/Zr ratios of various films processed by different methods. Serious Pb deficiency occurs in RTA-treated films, while FH- and FHC-processed films maintain stoichiometric Pb. Notably, RTA- and FH-treated films exhibit AFE characteristics, whereas FHC yields a RAFE state due to its sub-grain microstructure and associated nano-domains. Therefore, by comparing RTA- and FH-treated films, we can isolate the impact of Pb deficiency on the device performance; while by comparing FH- and FHC-treated films, we can estimate the contribution from changes in microstructure and domain scale. **Fig. S19** quantifies these effects through breakdown electric field $E_b$ (left) and energy storage density $U_e$ (right). Eliminating Pb loss enhances both $E_b$ and $U_e$, but microstructure optimization drives substantially greater improvement. We therefore conclude that microstructural engineering dominates energy storage enhancement.

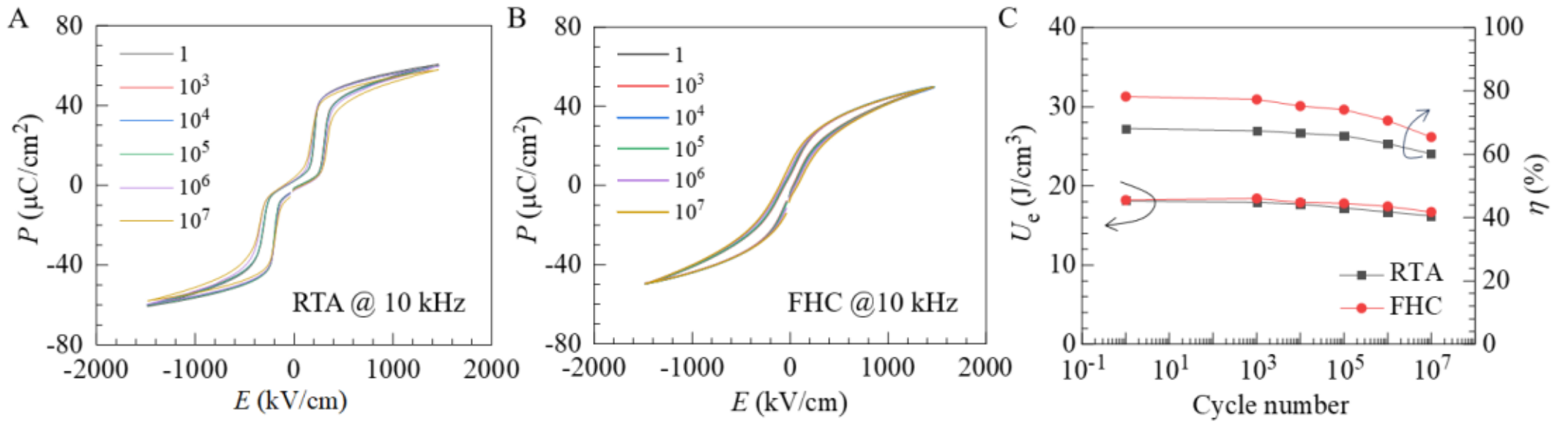

**Fig. S20. Fatigue performance of the PZO capacitors.** *P-E* loops at 10 kHz and room temperature during the fatigue measurement for (A) AFE-RTA film and (B) RAFE-FHC film. We employed a triangle waveform electric field of 1500 kV/cm at 500 kHz to ensure full polarization switching. C. The energy storage density ($U_e$, left) and efficiency ($\eta$, right) as a function of cycling numbers derived from (A) and (B). As shown in (A) and (B), both RTA and FHC devices exhibit minimal degradation in their *P-E* loops after $10^7$ cycles, demonstrating excellent energy storage stability (C). Specifically, the energy storage density ($U_e$) degraded by 10.5% (RTA) and 8.2% (FHC), and the efficiency ($\eta$) degraded by 11.7% (RTA) and 16.4% (FHC) after $10^7$ endurance cycles.

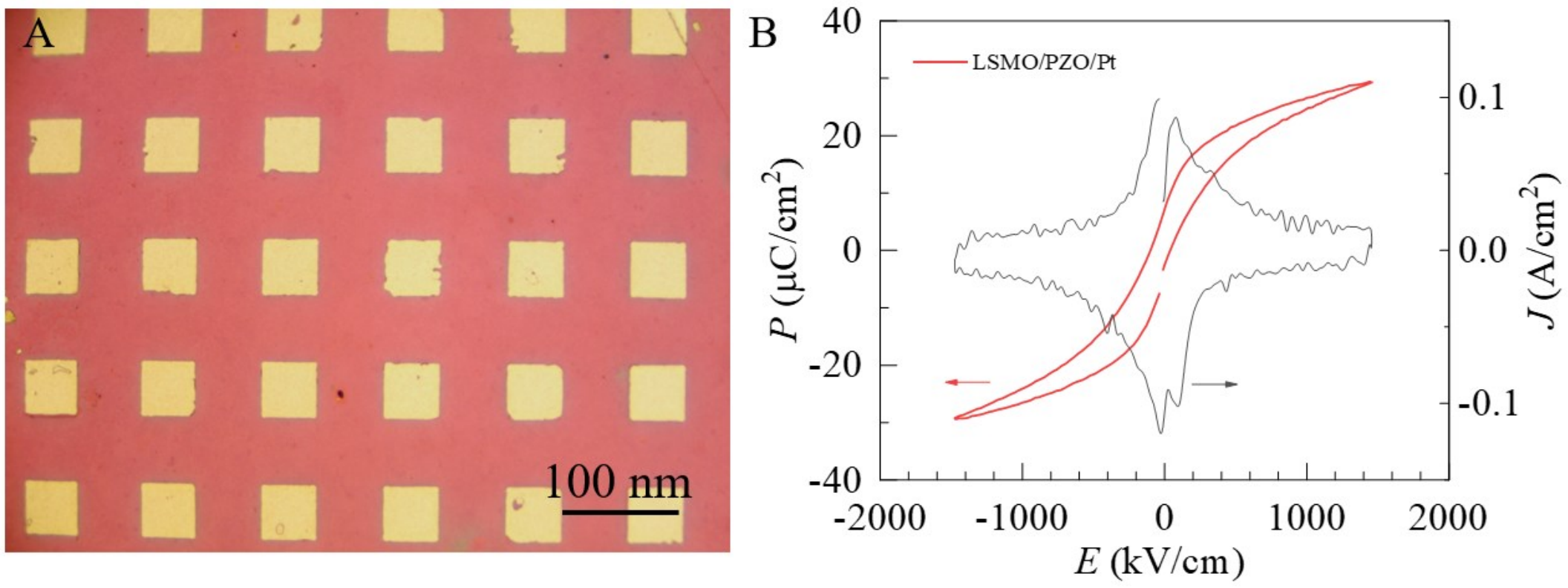

**Fig. S21. Ultrafast crystallization of the LSMO/PZO-300 nm/Pt capacitor device**. (A) Optical image of the devices after FHC process. (B) *P*-*E* hysteresis loop (left) and the corresponding switching current loop (right) of a representative device. Here we treat the entire capacitor devices (not solely the PZO film) directly by the FHC process. As shown in the optical image (A), the devices remain in good condition following FHC, and they exhibit relaxor antiferroelectric hysteresis loops as expected.

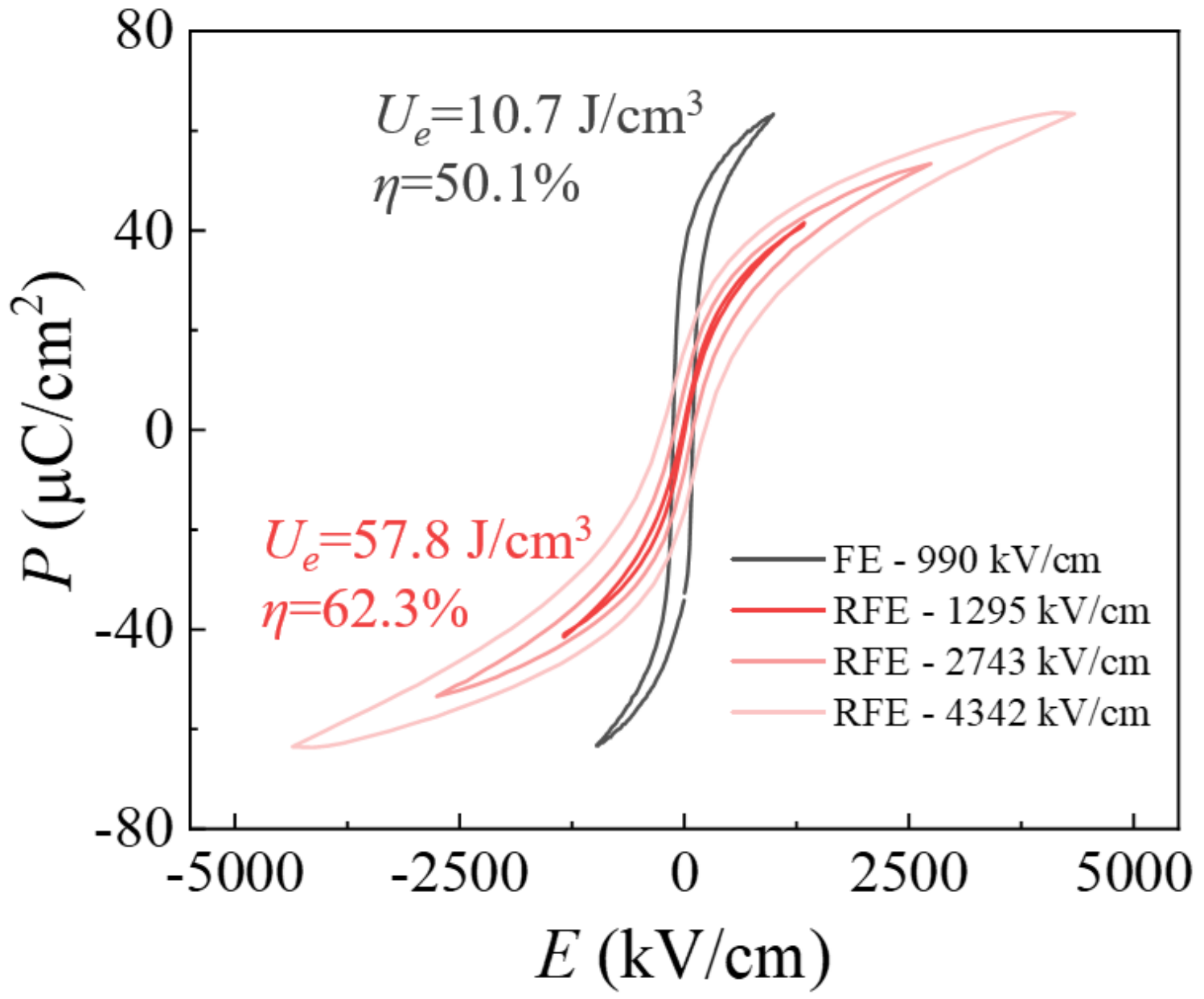

**Fig. S22. *P-E* loops of PZT films treated by RTA and FHC.** Utilizing the FHC method, we can transform the PZT film from the ferroelectric phase (FE) to the relaxor ferroelectric (RFE) phase. In conjunction with this phase transition, there is a marked enhancement in energy storage performance, with $U_e$ and $\eta$ increasing from 10.7 J/cm$^3$ and 50.1% for the ferroelectric phase to 57.8 J/cm$^3$ and 62.3% for the relaxor ferroelectric phase, respectively.

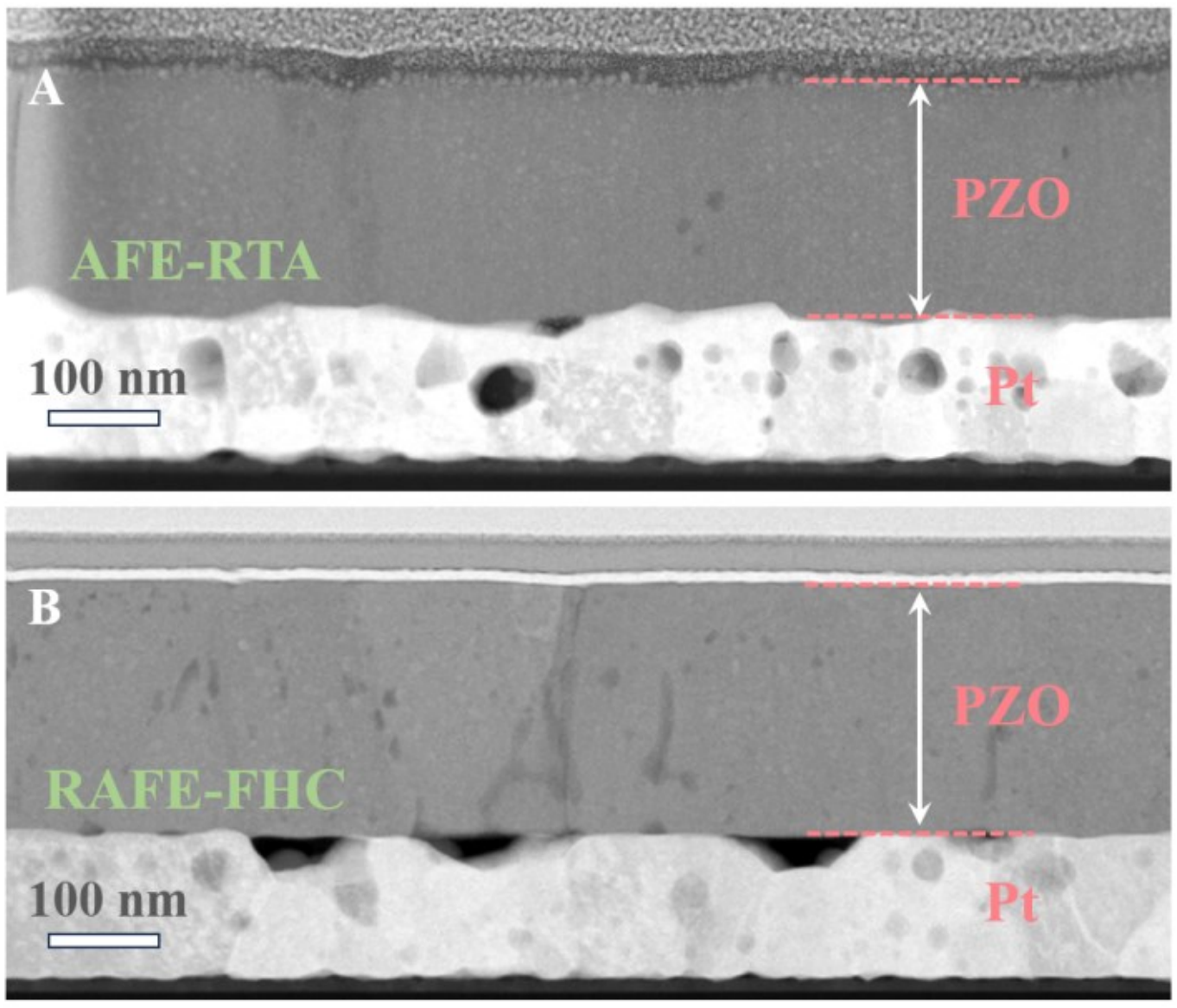

**Fig. S23. TEM cross-sectional images.** (**A**) AFE-RTA and (**B**) RAFE-FHC films. The films fabricated by both methods exhibit the same thickness of ~260 nm.

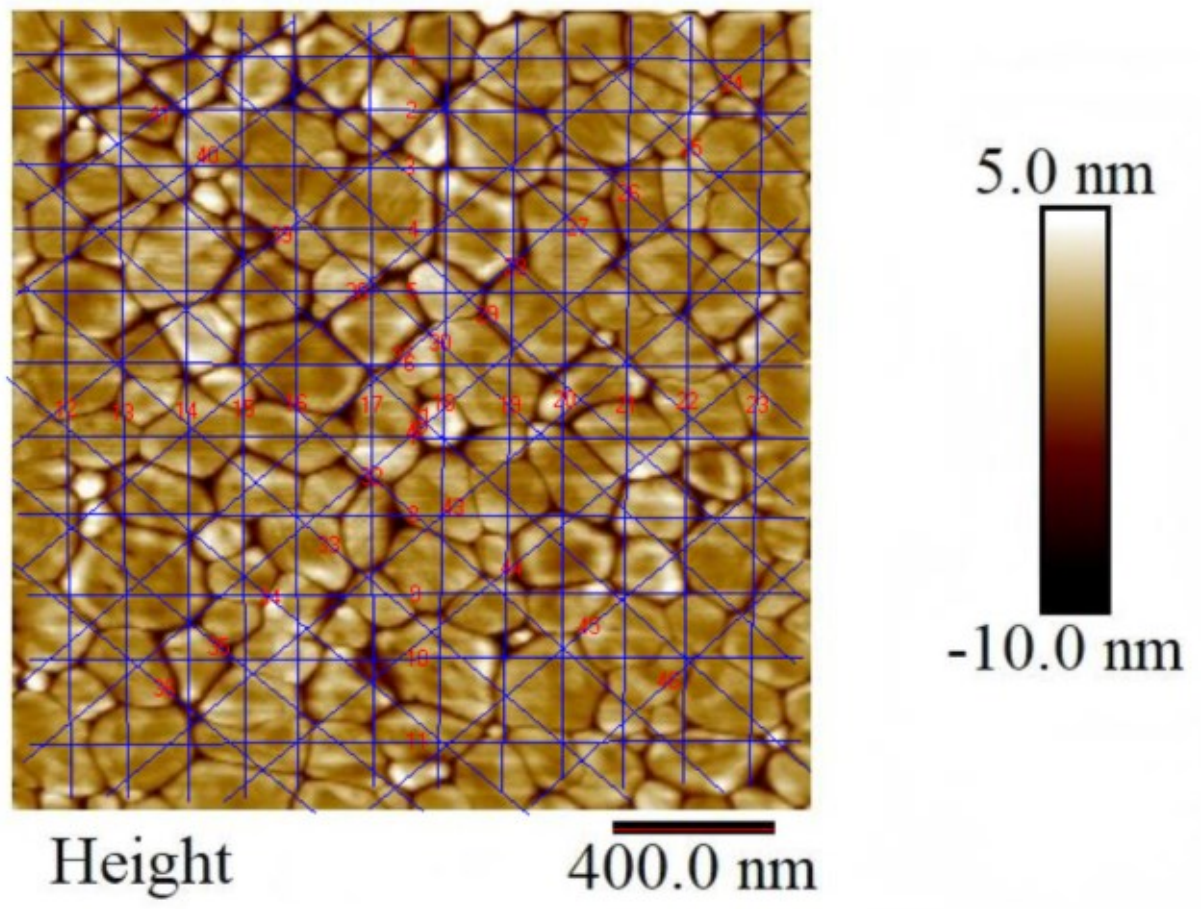

**Fig. S24. Method for grain size statistical analysis.** We analyzed grain size from AFM surface image using the intercept method (Heyn method) implemented in Nano Measurer software. This approach determines the average grain size ($\bar{d}$) by calculating the ratio of the total length of random test lines ($L$) to the number of grain boundary intersections ($N$), expressed as $\bar{d} = L/N$ . The derived mean intercept length corresponds to the effective grain diameter. For anisotropic microstructures, measurements included both horizontal and vertical test line orientations (as shown by blue lines in AFM image) to obtain a weighted average that accounts for grain morphology. The analysis yielded grain size distribution histograms and statistical parameters, providing quantitative characterization of both equiaxed and non-equiaxed grain structure.

**Table S1. Typical parameters used in various processing conditions.**

| Processing method | Heating rate (°C/s) | Heating time (s) | Temp. (°C) | Holding time (s) | Cooling method | Cooling time (s) |
|---|---|---|---|---|---|---|
| CA | 1 | 700 | 700 | 1800 | air | 1000 |
| RTA | 30 | 22 | 650 | 180 | air | 1000 |
| FH | 1000 | 0.65 | 650 | 0 | air | 20 |
| FHC | 1000 | 0.65 | 650 | 0 | liquid nitrogen | <1 |

**Table S2. Summary of bandgap ($E_g$) values from first-principles calculations and experimental measurements.** For first-principles calculations, we use the Perdew-Burke-Ernzerhof (PBE) functional method, and consider the effect of lattice parameters and Pb vacancies ($V_{Pb}$). The lattice parameters are taken from experimental results. It is found that RTA film with a $V_{Pb}$ concentration of 12.5% has a $E_g$ of 3.20 eV, which is ~0.6 eV larger than that of FHC film ($E_g$ ~ 2.61 eV). While experimentally, RTA film has a $E_g$ of 3.37 eV, which is ~0.3 eV larger than that of FHC film ($E_g$ ~ 3.06 eV). Therefore, PBE calculations successfully reproduce the experimental trend of bandgap reduction in FHC film relative to RTA film, despite systematically underestimating the bandgap values due to the known bandgap limitation of PBE functionals.

| Film | $a$ (Å) | $c$ (Å) | $V_{Pb}$ | $E_g$ cal. (eV) | $E_g$ exp. (eV) |
|---|---|---|---|---|---|
| AFE-RTA | 4.17 | 4.13 | No | 2.73 | |
| AFE-RTA | 4.17 | 4.13 | Yes, 12.5% | 3.20 | 3.37 |
| RAFE-FHC | 4.18 | 4.18 | No | 2.61 | 3.06 |